\documentstyle[seceq,supplement]{ptptex}
\begin{document}
\begin{center}
{\bf\Large Chapter 7}
\end{center}
\vskip 0.4cm
\begin{center}
{\bf\Large Gravitational Radiation Reaction}
\end{center}
\bigskip

\centerline{\large
Yasushi Mino$^{1,2}$, Misao Sasaki$^1$ and Takahiro Tanaka$^1$
}
\bigskip

\begin{center}
{\em $^1$Department of Earth and Space Science,
 Osaka University, Toyonaka 560, Japan}\\
{\em $^2$Department of Physics, 
Kyoto University, Kyoto 606, Japan}\\
\end{center}
\bigskip
\centerline{\large\bf Abstract}
\medskip

In this chapter, we consider the radiation reaction to the motion of a
point-like particle of mass $m$ and specific spin $S$ traveling on a
curved background. Assuming $S=O(Gm)$ and $Gm\ll L$ where $L$ is the
length scale of the background curvature, we divide the spacetime into
two regions; the external region where the metric is approximated by the
background metric plus perturbations due to a point-like particle and
the internal region where the metric is approximated by that of a black
hole plus perturbations due to the tidal effect of the background
curvature, and use the technique of the matched asymptotic expansion to
construct an approximate metric which is valid over the entire region.
In this way, we avoid the divergent self-gravity at the position of the
particle and derive the equations of motion from the consistency
condition of the matching. The matching is done to the order necessary
to include the effect of radiation reaction of $O(Gm)$ with respect to
the background metric as well as the effect of spin-induced force. The
reaction term of $O(Gm)$ is found to be completely due to tails of
radiation, that is, due to curvature scattering of gravitational
waves. In other words, the reaction force is found to depend on the
entire history of the particle trajectory. Defining a regularized metric
which consists of the background metric plus the tail part of the
perturbed metric, we find the equations of motion reduce to the geodesic
equation on this regularized metric, except for the spin-induced force
which is locally expressed in terms of the curvature and spin tensors.
Some implications of the result and future issues are briefly discussed.

\section{Introduction}
\label{sec:intro}

The problem of radiation reaction has long been 
one of the fundamental theoretical issues in general relativity. 
Starting from the historical works of Eddington 
in his 1922 book\cite{Eddington}, 
Chandrasekhar and Esposito\cite{ChandraEspo} discussed 
the radiation reaction of the self-gravitating fluid 
emphasizing the importance of the time asymmetric part of the metric 
appearing in the post-Newtonian expansion, 
and Burke and Thorne\cite{BurkeThorne} found that 
the leading contribution from the time asymmetric part 
can be compactly expressed in the form of 
a resistive potential. 

The previous studies of radiation reaction\cite{ChandraEspo,BurkeThorne} 
were done under the assumption 
that the post-Newtonian expansion is valid. 
Here we consider this problem 
in the framework of linear perturbation theory in a general spacetime.
A part of motivation is to give a rigid foundation 
of the method to solve the Einstein equations perturbatively as an
expansion with respect to the perturbation caused by a point-like
particle. Usually one adopts a point-like particle to 
represent a black hole or neutron star 
in the linear perturbation studies as was done in Chapter 1. 
Then one may pose several questions:
Since the perturbed field diverges at the location 
of the point-like particle, 
is the approximation scheme of linear perturbation still valid? 
Does the point-like particle really represent 
a black hole or a neutron star? 
If it represents a black hole, the center of it is 
inside the event horizon, and then in what sense 
does `the motion of the particle' make sense? 
Here we are going to clarify the meaning of 
the particle trajectory and derive the equations of motion including
the effect of radiation reaction to the first non-trivial order. 

Before starting the discussion of the gravitational 
radiation reaction, it is worthwhile to refer to the electromagnetic
case in a fixed curved background spacetime which was discussed by
DeWitt and Brehme\cite{DeWitt}.
In the electromagnetic case, the total energy momentum tensor 
composed of the particle and field 
contributions satisfies the conservation law. 
The conservation law is integrated over the interior of a world tube
with an infinitesimal length surrounding the particle orbit.
The part of the integration which does not vanish in the limit of
small tube radius is transformed into 
the surface integrations over both ends of the tube 
and over the surface of the tube by using the Gauss theorem. 
The integrations over the top and bottom of the tube, respectively, 
give the definition of the particle momenta at both ends and 
the difference between them represents the change of the momentum 
during this infinitesimal time interval, which is to be equated 
with the momentum flow given by the 
integration over the surface of the tube. 
In this way the equations of motion are obtained. 

In the case of gravitational radiation reaction, 
it is possible to construct a conserved rank-two tensor 
defined on the background spacetime, 
composed of the matter field and the metric perturbation\cite{Mino1}. 
However, there is an essential difference between the electromagnetic 
and gravitational cases. 
In electromagnetism, we can consider an extended 
charge distribution which is supported by a certain force other 
than the electromagnetic field. 
Thus it is possible to assume that the charge and mass 
distributions of a point-like particle are not distorted by 
the effect of the radiation reaction. 
Therefore one may consistently assume that 
the momentum and the electric current of the particle are 
proportional to the 4-velocity of the particle. 
Moreover the electromagnetic charge $e$ is not directly related 
to the energy momentum of the particle which is proportional 
to the mass $m$. 
Hence, even if the limit of zero particle radius 
is taken, the divergent self-energy ($\propto e^2$) 
can be renormalized into the mass.
In the case of gravitational radiation reaction, 
it is not possible to consider such an ideal point-like 
particle because every force field universally couples with gravity. 
Even worse, the role of $e$ in electromagnetism is also attributed to
$m$. Thus a simple renormalization scheme does not make any sense. 

In order to deal with the gravitational case, 
we use the matched asymptotic expansion technique 
that has been studied by many authors (e.g.,
D'Eath \cite{Death} and Thorne and Hartle \cite{Thorne1}) 
in the context of the post-Minkowski (or post-Newtonian) 
approximation. We assume that the metric sufficiently far from the
particle is approximated by the perturbation on the background spacetime
generated by a point-like particle. We call this the external metric.
We also assume that the internal metric which describes the 
geometry around the particle is represented by 
a black hole metric of small mass in the lowest order approximation.
As the particle moves in the curved background, the internal metric
suffers from the tidal distortion. 
Thus both internal and external metrics are constructed perturbatively.
The expansion parameters for the internal and external metrics are,
however, different. We call this construction of the metric in the
internal region the internal scheme and that in the external region the
external scheme. 
Assuming the existence of the matching region where
both schemes are valid, the internal and external metrics are expanded
there as double series with respect to the two expansion parameters.
Then the terms in these series are labeled by two indices 
which denote the powers of the two expansion parameters. 
Equating them order by order, we obtain the matching condition, 
through which one scheme determines the 
boundary condition of the other and vice versa. 

Using the matched asymptotic expansion to the first non-trivial orders
of the expansion parameters, we present two different derivations 
of the equations of motion with the radiation reaction force of O$(Gm)$; 
(1) by means of an explicit construction of 
the metric, and (2) by using the so-called laws of motion and
precession\cite{Thorne1}.

As mentioned above, in constructing the internal metric, 
the tidal distortion of the geometry is taken into account 
by the perturbation of the black hole. 
In the method (1), we set the gauge condition 
in the internal metric so that $J=0$ and $1$ linear homogeneous
perturbations of the black hole vanish since 
they are purely gauge degrees of freedom as long as both the mass and
angular momentum of the black hole stay constant. 
Applying a limited class of coordinate 
transformations that keep the meaning 
of the center of the particle unambiguous, 
the internal metric is matched with 
the external one in the matching region. 
Then we find that for a given trajectory of the particle 
a consistent coordinate transformation does not always exist, 
and this consistency condition determines 
the equations of motion. 

In the method (2), not all the metric components are evaluated 
in both schemes independently but we assume 
the existence of a coordinate transformation 
that gives a relation between the internal and external metrics. 
Once we know some metric components in one scheme, 
the counter parts in the other scheme are obtained 
from the matching condition. At this stage, the gauge condition is not
fixed in a unique manner.
The coordinate transformation between the internal metric and 
the external metric is chosen so that some of the metric components
that are evaluated in both schemes are correctly matched in the matching
region. Substituting the metric constructed in this way 
into the Einstein equations, we obtain the consistency 
condition. There is a convenient method to extract out the information
about the equations of motion from the consistency condition. 
Namely to use the laws of motion and precession
introduced by Thorne and Hartle\cite{Thorne1}. 
The laws of motion and precession are derived from
the non-covariant but conserved form of the Einstein equations.

The resulting equations obtained from both derivations 
are the same, although the strategies are quite different. 
In the method (1), the metrics in both schemes are 
calculated independently by using the Einstein equations. 
The matching condition is used to obtain the consistency conditions,
which in turn give the equations of motion. 
On the other hand, in the method (2), the matching condition 
is used to construct the metric. The consistency condition is
derived by requiring that thus obtained metric satisfies 
the Einstein equations. 
The meaning of the matching condition in deriving the equations of
motion is clearer in the method (1) than in (2), but the method (2) is
much simpler and straightforward than the method (1) as we shall see
in the following.

The organization of this chapter is as follows.
We use the terminology `a monopole (spinning) particle' 
to refer to a particle which represents a Schwarzschild (Kerr) black
hole. In section \ref{sec:MAE}, the matched asymptotic 
expansion technique is explained in detail. 
In section \ref{sec:minoExt}, we discuss the metric 
perturbation in the external scheme. 
In section \ref{sec:deri1}, the equations of motion for a monopole
particle are derived by using the method (1).
The method (1) is applied only to the case of 
a monopole particle because of the difficulty in constructing
the perturbed metric of a Kerr black hole.
The case for a spinning particle is considered 
in section \ref{sec:deri2} by using the method (2).  

Throughout this chapter we assume that the background metric satisfies
the vacuum Einstein equations\footnote{The result is not altered 
even if we assume that the background spacetime is 
vacuum just around the particle.}.
Hence in the following calculations we use the fact that 
the background Ricci tensor vanishes;
\begin{eqnarray}
R_{\mu\nu}=0.
\end{eqnarray}

\section{Matched Asymptotic Expansion}
\label{sec:MAE}

The matched asymptotic expansion is a technique 
with which the same physical quantities derived 
in different zones by two different approximation schemes
are matched in the overlapping region 
to obtain an approximate solution valid in the whole region. 
We first prepare the metrics in both internal and external zones 
by using different approximation schemes. 
The internal zone is the region 
where the self-gravity of the particle dominates 
while the external zone is the region where 
the background geometry dominates the full geometry. 

In the internal zone, we assume that the metric can be described 
by that of a black hole plus perturbation. 
Namely, we assume that the particle is represented by 
a Schwarzschild/Kerr black hole in the lowest order of 
approximation.
In the present case, the perturbation 
is caused by the tidal effect of the curvature of the 
spacetime in which the particle travels. As mentioned in Introduction,
we call this construction of the metric the internal scheme. 
In order to make this scheme valid, the linear extension of
the internal zone around the particle 
must be much smaller than the background curvature scale $L$. 
We introduce the coordinate 
$\{ X^a \}=\{ T, X^i \}\quad (a=0,1,2,3;~i=1,2,3)$ 
for the internal scheme and $|X|(:=\sqrt{X^iX^i})$ is assumed to
represent the physical distance scale\footnote{
In this chapter, we adopt the Minkowskian summation rule on
$a,b,\cdots$, and the Kronecker summation rule on $i,j,\cdots$ 
over the repeated indices.}.
Then the internal scheme is valid when
\begin{eqnarray}
|X| \ll L\,,
\end{eqnarray}
where $L$ is the length scale of the background curvature. 

In the external zone, 
we expect that the metric is well approximated by the perturbation 
induced by a point source on a given background spacetime. 
We call this construction of the metric the external scheme. 
This approximation scheme is valid 
when the self-gravity of the particle is sufficiently weak, 
that is, 
\begin{eqnarray}
Gm \ll |X| \,,
\end{eqnarray}
where $(Gm)$ is the scale of Schwarzschild radius. 
As the point source is placed where the external scheme is invalid, 
there is no matter source in the external zone. 
Thus the external metric is given by a vacuum solution of the Einstein
equations. 

We require that the metrics obtained in both schemes be matched 
in the overlapping region of both zones, 
by considering a coordinate transformation 
between the internal and external metrics. 
Safely, we may assume the existence of the matching region 
as long as 
\begin{eqnarray}
Gm \ll L\,, 
\end{eqnarray}
is satisfied. For definiteness, we set the matching radius at 
\begin{eqnarray}
|X| \sim(GmL)^{1/2},
\end{eqnarray}
in the spatial coordinates of the internal scheme, {$X^i$}. 
Then writing down the metric in the internal scheme, 
we have two independent small parameters $|X|/L$ and $Gm/|X|$ 
in the matching region. 
The power expansion with respect to these two small parameters 
allows us to consider the matching order by order. 

First we consider the expansion of the internal scheme. 
Recalling that the perturbation in the internal zone is 
induced by the external curvature 
which has a characteristic length scale $L$, 
the metric can be expanded in powers of $|X|/L$ as 
\begin{equation}
\tilde g_{ab}(X) = 
{}^{(0)}H_{ab}(X)+{1\over L}{}^{(1)}H_{ab}(X)
+{1\over L^2}{}^{(2)}H_{ab}(X)+\cdots, 
\label{eq:bh0}
\end{equation}
where ${}^{(0)}H_{ab}(X)$ is the unperturbed black hole metric.
We expect that ${}^{(1)}H_{ab}(X)$ will be given by the standard linear
perturbation of the black hole. Later, we find that ${}^{(1)}H_{ab}(X)$
can be consistently set to zero, which is in accordance with the notion
that the spacetime curvature is of $O(1/L^2)$. Thus the standard black
hole perturbation theory applies up to ${}^{(2)}H_{ab}(X)$. 
Further we expand the metric with respect to $Gm/|X|$ 
which is also small at the matching radius: 
\begin{eqnarray}
{}^{(0)}H_{ab}(X)&=&\eta_{ab}+Gm{}_{(1)}^{(0)}H_{ab}(X)
+(Gm)^2 {}_{(2)}^{(0)}H_{ab}(X) 
+\cdots\,,
\nonumber \\ 
{1\over L}{}^{(1)}H_{ab}(X)&=&
{1\over L}{}_{(0)}^{(1)}H_{ab}(X)
+{Gm\over L}{}_{(1)}^{(1)}H_{ab}(X)
+{(Gm)^2\over L}{}_{(2)}^{(1)}H_{ab}(X)+\cdots\,,
\nonumber \\ 
{1\over L^2}{}^{(2)}H_{ab}(X)&=&
{1\over L^2}{}_{(0)}^{(2)}H_{ab}(X)
+{Gm\over L^2}{}_{(1)}^{(2)}H_{ab}(X)
+{(Gm)^2\over L^2}{}_{(2)}^{(2)}H_{ab}(X)+\cdots\,.
\label{eq:bh}
\end{eqnarray}
Note that, from the definitions of the expansion parameters,
the ${}_{(n)}^{(m)}H_{ab}$ component of the metric behaves as 
\begin{eqnarray}
{}_{(n)}^{(m)}H_{ab} \sim |X|^{m-n}. 
\label{eq:bhpower} 
\end{eqnarray}

The explicit form of the coordinate transformation 
from the general coordinates of a background metric $\{ x^\mu \}$ 
to the coordinates of the internal scheme $\{X^a\}$ 
will be discussed in section \ref{sec:minoExt} for the method (1)
and in section \ref{sec:deri2} for the method (2). 
Assuming the matching can be consistently done, the full metric 
in the external scheme $\tilde g_{\mu\nu}(x)$ 
is written in terms of the internal coordinates as
\begin{equation}
\tilde g_{ab}(X)dX^a dX^b=\tilde g_{\mu\nu}(x)dx^{\mu} dx^{\nu} \,.
\label{eq:465}
\end{equation}

Generally, as the external metric can be expanded by $Gm/|X|$, 
we write it as 
\begin{eqnarray}
 \tilde g_{ab}(X)= g_{ab}(X) +Gm {}_{(1)}h_{ab}(X)
+(Gm)^2 {}_{(2)}h_{ab}(X) +\cdots. 
\end{eqnarray}
Then $Gm {}_{(1)}h_{ab}(X)$ can be recognized as 
the linear perturbation on the background $g_{ab}(X)$. 
Further we expand it with respect to $|X|/L$ as
\begin{eqnarray}
g_{ab}(X) &=& 
{}^{(0)}_{(0)}h_{ab}(X)+{1\over L}{}^{(1)}_{(0)}h_{ab}(X)
+{1\over L^2}{}^{(2)}_{(0)}h_{ab}(X)+\cdots\,,
\nonumber \\
Gm {}_{(1)}h_{ab}(X) &=& 
Gm{}_{(1)}^{(0)}h_{ab}(X)+{Gm\over L}{}_{(1)}^{(1)}h_{ab}(X)
+{Gm\over L^2}{}_{(1)}^{(2)}h_{ab}(X)+\cdots\,,
\nonumber \\ 
(Gm)^2 {}_{(2)}h_{ab}(X) &=& 
(Gm)^2{}_{(2)}^{(0)}h_{ab}(X)+{(Gm)^2\over L}
{}_{(2)}^{(1)}h_{ab}(X)
\nonumber\\
&&\qquad+{(Gm)^2\over L^2}{}_{(2)}^{(2)}h_{ab}(X) +\cdots\,. 
\label{eq:ext}
\end{eqnarray}
As before, 
\begin{eqnarray}
{}_{(n)}^{(m)}h_{ab} \sim |X|^{m-n}. 
\label{eq:extpower}
\end{eqnarray}

For brevity, we call ${}_{(n)}^{(m)}h_{ab}$ or 
${}_{(n)}^{(m)}H_{ab}$ the $({}^{m}_{n})$-component and 
the matching condition for them as the $({}^{m}_{n})$ matching. 
In the matching region ($|X| \sim (GmL)^{1/2}$), the 
$({}^{m}_{n})$-component is of $O\left((Gm/L)^{(m+n)/2}\right)$. 
The matching condition requires that all the corresponding terms 
in Eqs.~(\ref{eq:bh}) and (\ref{eq:ext}) should be identical. 
Then the matching condition is given by 
equating the terms of the same power 
in $|X|$ in both schemes to desired accuracy.
Thus the condition for the $({}^{m}_{n})$ matching is 
\begin{equation}
 \sum_{m'-n'=m-n \atop m'\le m} {(Gm)^{n'} \over L^{m'} } 
 {}_{(n')}^{(m')}h_{ab} = 
 \sum_{m'-n'=m-n \atop m'\le m} {(Gm)^{n'} \over L^{m'} }
 {}_{(n')}^{(m')}H_{ab}
  +O\left({(Gm)^{n+1} \over L^{m+1} }|X|^{(m-n)}\right). 
\end{equation}

\section{External Scheme}
\label{sec:minoExt}
As we assume that the gravitational radius of the particle, $Gm$, is
small compared with the length scale of the background curvature, $L$,
we approximate $\delta g_{\mu\nu}$ 
by the linear perturbation induced by a point-like particle, 
$h_{\mu\nu}$, in the whole spacetime region 
except for the vicinity of the world line of the particle. 
The calculation is performed in an analogous manner 
to the case of the scalar and vector perturbations 
developed by DeWitt and Brehme\cite{DeWitt}.

We take a Green function approach to 
study the linear perturbation of the metric generated by a point
source. In order to calculate the tensor Green function 
in a background covariant manner, 
we begin with introducing the concept of bi-tensors.


\subsection{Bi-tensor formalism}

Bi-tensors are tensors which depend on 
two distinct spacetime points, say, $x$ and $z$, 
so that they can have two types of indices.
The simplest example is given by a direct product of 
tensors at the points $x$ and $z$ as
\begin{eqnarray}
A^{\mu\alpha}(x,z)=B^\mu(x)C^\alpha(z) \,.
\end{eqnarray}
In what follows, we use $x$ for a field point and $z$ for a point on 
the particle trajectory, and assign the letters $\alpha$, $\beta$,
$\gamma$, $\delta$, $\epsilon$, $\zeta$, $\eta$ for the tensor indices
of $z$ and $\mu$, $\nu$, $\xi$, $\rho$, $\sigma$ for $x$. 

Basic bi-tensors used in our calculations are
half the squared geodetic interval $\sigma (x,z)$,
\begin{eqnarray}
&&\sigma(x,z)={1\over 2} g^{\mu\nu}(x)\sigma_{;\mu}(x,z)\sigma_{;\nu}(x,z)=
{1\over 2} g^{\alpha\beta}(z)\sigma_{;\alpha}(x,z)\sigma_{;\beta}(x,z) \,,
\nonumber \\ 
&& \lim_{x\rightarrow z}\sigma (x,z)_{;\mu} = 
\lim_{x\rightarrow z}\sigma (x,z)_{;\alpha} = 0 \,,
\label{111db}
\end{eqnarray}
and the geodetic parallel displacement bi-vector,
\begin{eqnarray}
&&\bar g_{\mu\alpha;\nu}(x,z)g^{\nu\sigma}(x) \sigma_{;\sigma}(x,z) =0,
\quad
\bar g_{\mu\alpha;\beta}(x,z)g^{\beta\gamma}(z) 
\sigma_{;\gamma}(x,z) =0,
\nonumber\\
&&\lim_{x\rightarrow z}\bar g_{\mu}{}^{\alpha} (x,z) 
= \delta_{\mu}{}^{\alpha}.
\label{131db}
\end{eqnarray}
These are used to expand bi-tensors around the orbit of a particle. 
For example, we have 
\begin{eqnarray}
 A^\alpha(x,z) &=& \lim_{x'\rightarrow z}
\biggl(A^\alpha(x',z)-\sigma_{;\mu'}(x,x')A^{\alpha;\mu'}(x',z)
+O(\epsilon^2)\biggr) \,,
\\ 
 B^\mu(x) &=& \bar g^\mu{}_\alpha(x,z)
 \biggl(B^\alpha(z)-\sigma_{;\beta}(x,z)B^{\alpha;\beta}(z)
+O(\epsilon^2)\biggr) \,,
\end{eqnarray} 
for a small geodetic interval between $x$ and $z$, 
where $\epsilon = \sqrt{2|\sigma(x,z)|}$.
These relations can be verified by taking the $x\to z$ limit 
of their repeated derivatives. 

By evaluating the repeated derivatives of Eqs.~(\ref{111db}) and 
(\ref{131db}) in the coincidence limit $x\rightarrow z$,
we obtain some useful formulas for expansion in $\epsilon$:
\begin{eqnarray}
\sigma_{;\alpha\beta}(x,z) &=& g_{\alpha\beta}(z)- {1\over 3} 
R_{\alpha}{}^{\gamma}{}_{\beta}{}^{\delta}(z) 
\sigma_{;\gamma}(x,z) \sigma_{;\delta}(x,z) +O(\epsilon^3) \,, 
\label{128db} \\ 
\sigma_{;\mu\beta}(x,z) &=& -\bar g_{\mu}{}^{\alpha}(x,z)
\left(g_{\alpha\beta}(z)+{1\over 6} 
R_{\alpha\gamma\beta\delta}(z) 
\sigma^{;\gamma}(x,z) \sigma^{;\delta}(x,z)\right)
\cr &&
+O(\epsilon^3) \,,
\label{173db} \\ 
\bar g^{\mu\alpha}{}_{;\beta}(x,z) & = & -{1\over 2} 
\bar g^{\mu\gamma}(x,z)R^{\alpha}{}_{\gamma\beta\delta}(z) 
\sigma^{;\delta}(x,z) +O(\epsilon^2) \,, \cr
\bar g^{\mu\alpha}{}_{;\nu}(x,z) & = & -{1\over 2} 
\bar g^{\mu\beta}(x,z) \bar g_{\nu}{}^{\gamma}(x,z) 
R^{\alpha}{}_{\beta\gamma\delta}(z) 
\sigma^{;\delta}(x,z) +O(\epsilon^2) \,. 
\label{140db}
\end{eqnarray}
We also introduce the van Vleck-Morette determinant, $\Delta(x,z)$: 
\begin{equation}
\Delta(x,z):=|\bar g^{\alpha\mu}(z,x)\sigma_{;\mu\beta}(x,z)|, 
\label{Deltadef}
\end{equation}
which appears in the expression of the Green function later.


\subsection{Tensor Green function}

We consider the linearized Einstein equations. 
We introduce the trace-reversed metric perturbation, 
\begin{eqnarray}
\psi_{\mu\nu}(x) &=& h_{\mu\nu}(x) -{1\over 2}g_{\mu\nu}(x) h(x) \,, 
\end{eqnarray}
and set the harmonic gauge condition,
\begin{eqnarray}
\psi^{\mu\nu}{}_{;\nu}(x)&=&0 \,,
\end{eqnarray}
where $h(x)$ and $\psi(x)$ are 
the trace of $h^{\mu\nu}(x)$ and that of $\psi^{\mu\nu}(x)$, respectively, 
and the semicolon means the covariant derivative with respect to the 
background metric. In this gauge, the linearized Einstein equations become
\begin{eqnarray}
-{1\over 2}\psi^{\mu\nu;\xi}{}_\xi(x)
-R^\mu{}_\xi{}^\nu{}_\rho (x) \psi^{\xi\rho}(x)
= 8 \pi G T^{\mu\nu}(x) \,.
\end{eqnarray}
Thus we define the tensor Green function $G^{\mu\nu\alpha\beta}(x,z)$ 
which satisfies 
\begin{eqnarray}
G^{\mu\nu\alpha\beta;\xi}{}_{;\xi}(x,z)
&&+2R^\mu{}_\xi{}^\nu{}_\rho(x) G^{\xi\rho\alpha\beta}(x,z) 
\cr
&&=-2\bar g^{\alpha(\mu}(x,z)\bar g^{\nu)\beta}(x,z)
{\delta^{(4)}(z-x) \over \sqrt{-g}} \,, 
\label{eq:green}
\end{eqnarray}
where $g$ is the determinant of the metric $g_{\mu\nu}(x)$. 

First we consider 
the elementary solution $G_{*}^{\mu\nu\alpha\beta}(x,z)$ which satisfies 
Eq.~(\ref{eq:green}) except at the $\sigma(x,z)\rightarrow 0$ limit and 
takes the Hadamard form, 
\begin{eqnarray}
G_*^{\mu\nu\alpha\beta}(x,z)=
{1\over (2\pi)^2}\Biggl({u^{\mu\nu\alpha\beta}(x,z)\over\sigma(x,z)}
&&+v^{\mu\nu\alpha\beta}(x,z)\log|\sigma(x,z)|
\cr &&
+w^{\mu\nu\alpha\beta}(x,z)\Biggr) \,. 
\label{eq:Hadamard}
\end{eqnarray}
The bi-tensors $u^{\mu\nu\alpha\beta}(x,z)$, $v^{\mu\nu\alpha\beta}(x,z)$ 
and $w^{\mu\nu\alpha\beta}(x,z)$ are regular in the
$\sigma(x,z)\rightarrow 0$ limit and $u^{\mu\nu\alpha\beta}(x,z)$
satisfies the normalization condition, 
\begin{eqnarray}
\lim_{x\rightarrow z}u^{\mu\nu\alpha\beta}(x,z)= 
\lim_{x\rightarrow z}2\bar g^{\alpha(\mu}(x,z)\bar g^{\nu)\beta}(x,z) \,. 
\label{eq:unorm} 
\end{eqnarray}
If we put the form (\ref{eq:Hadamard}) into the left hand side of 
Eq.~(\ref{eq:green}), the terms can be classified into three parts.  
One is the terms which contain the factor $1/\sigma^2(x,z)$ manifestly 
and another is the terms which contain $\log|\sigma(x,z)|$. 
The remaining terms have no singular behavior at the
$\sigma(x,z)\rightarrow 0$ limit. 
Since the form (\ref{eq:Hadamard}) is redundant, 
we can set these three sets to vanish separately:
\begin{eqnarray} 
&& \left(2u^{\mu\nu\alpha\beta;\xi}(x,z) 
-{\Delta^{;\xi}(x,z)\over\Delta(x,z)}u^{\mu\nu\alpha\beta}(x,z)\right) 
\sigma_{;\xi}(x,z) = 0 \,, 
\label{eq:ueq}
\\ 
&& v^{\mu\nu\alpha\beta;\xi}{}_{;\xi}(x,z) 
+ 2 R^\mu{}_\xi{}^\nu{}_\rho(x) v^{\xi\rho\alpha\beta}(x,z) = 0 \,, 
\label{eq:veq}
\\ 
&& 2v^{\mu\nu\alpha\beta}(x,z) 
+\left(2v^{\mu\nu\alpha\beta;\xi}(x,z) 
-{\Delta^{;\xi}(x,z)\over\Delta(x,z)}v^{\mu\nu\alpha\beta}(x,z)\right) 
\sigma_{;\xi}(x,z) 
\nonumber \\ 
&& \qquad \quad 
+u^{\mu\nu\alpha\beta;\xi}{}_{;\xi}(x,z) 
+2 R^\mu{}_\xi{}^\nu{}_\rho(x) u^{\xi\rho\alpha\beta}(x,z) 
\nonumber \\ 
&& \qquad \quad 
+\left(w^{\mu\nu\alpha\beta;\xi}{}_{;\xi}(x,z)
+2 R^\mu{}_\xi{}^\nu{}_\rho(x) w^{\xi\rho\alpha\beta}(x,z)\right)\sigma(x,z)
 = 0 \,. 
\label{eq:weq} 
\end{eqnarray}
Equation~(\ref{eq:ueq}) is solved   
with the normalization (\ref{eq:unorm}) as 
\begin{eqnarray} 
u^{\mu\nu\alpha\beta}(x,z)=
2\bar g^{\alpha(\mu}(x,z)\bar g^{\nu)\beta}(x,z)\sqrt{\Delta(x,z)} \,.
\label{eq:u}
\end{eqnarray}
The bi-tensors $v^{\mu\nu\alpha\beta}(x,z)$ and
$w^{\mu\nu\alpha\beta}(x,z)$ 
are to be determined by solving Eqs.~(\ref{eq:veq}) and (\ref{eq:weq}). 
The bi-tensor $w^{\mu\nu\alpha\beta}(x,z)$ is not needed but 
the bi-tensor $v^{\mu\nu\alpha\beta}(x,z)$ plays an important role in 
the following discussion. 
Although it is difficult to find 
the solution of $v^{\mu\nu\alpha\beta}(x,z)$ in an arbitrary 
background spacetime, 
its explicit form is not required for the succeeding discussions. 
However it is important to note that $v^{\mu\nu\alpha\beta}(x,z)$ is uniquely
determined. The reason is as follows.
{}From Eq.~(\ref{eq:veq}) one finds it satisfies a hyperbolic 
equation. Hence the problem is if its Cauchy data are unique or not.
First we note the coincidence limit of Eq.~(\ref{eq:weq}), which gives
\begin{eqnarray}
\lim_{x\rightarrow z}v^{\mu\nu\alpha\beta}(x,z)
=\lim_{x\rightarrow z}2\bar g^\alpha{}_{(\xi} (z,x)
\bar g^{\beta}{}_{\rho)}(z,x)
R^{\mu\xi\nu\rho}(x). 
\label{eq:ulim}
\end{eqnarray}
Then taking the null limit $\sigma(x,z)\rightarrow 0$ of
 Eq.~(\ref{eq:weq}), 
we obtain the first order differential equation for
$v^{\mu\nu\alpha\beta}(x,z)$ which can be solved along a null geodesic. 
Thus this equation with the boundary condition (\ref{eq:ulim}) 
uniquely determines $v^{\mu\nu\alpha\beta}(x,z)$ on the light cone 
emanating from $z$. 
Therefore the hyperbolic equation (\ref{eq:veq}) has a unique
solution.
We also mention that $v^{\mu\nu\alpha\beta}(x,z)$ is divergence free,
\begin{equation}
v^{\mu\nu\alpha\beta}{}_{;\nu}(x,z)=0. 
\label{divv}
\end{equation}
To see this we note the harmonic gauge condition on the Green 
function requires 
\begin{equation}
 \lim_{\sigma\rightarrow 0} v^{\mu\nu\alpha\beta}{}_{;\nu}(x,z)=0.
\end{equation}
We also see that the equation for  
$v^{\mu\nu\alpha\beta}{}_{;\nu}(x,z)$ follows from Eq.~(\ref{eq:veq}),
\begin{equation}
 \left[v^{\mu\nu\alpha\beta}{}_{;\nu}(x,z)\right]{}^{;\xi}{}_{;\xi}=0, 
\end{equation}
where we have used the fact $R^{\mu\xi\nu\rho}{}_{;\rho}=0$, 
which is proved by contracting the Bianchi identities 
for the vacuum case. Thus we conclude
that Eq.~(\ref{divv}) holds everywhere.

The Feynman propagator $G_F^{\mu\nu\alpha\beta}(x,z)$ can be derived 
from the elementary solution $G_*^{\mu\nu\alpha\beta}(x,z)$ 
by the $i\epsilon$-prescription. 
\begin{eqnarray}
G_F^{\mu\nu\alpha\beta}(x,z)
={1\over (2\pi)^2}&&
\Biggl({u^{\mu\nu\alpha\beta}(x,z)\over\sigma(x,z)+i\epsilon}
\cr&&
+v^{\mu\nu\alpha\beta}(x,z)\log(\sigma(x,z)+i\epsilon)
+w^{\mu\nu\alpha\beta}(x,z)\Biggr). 
\end{eqnarray}
The imaginary part of the Feynman propagator $G_F^{\mu\nu\alpha\beta}(x,z)$ 
gives the symmetric Green function $\bar G^{\mu\nu\alpha\beta}(x,z)$, 
from which we can obtain 
the retarded Green function $G_{Ret}^{\mu\nu\alpha\beta}(x,z)$, 
and the advanced Green function $G_{Adv}^{\mu\nu\alpha\beta}(x,z)$ as
\begin{eqnarray}
\bar G^{\mu\nu\alpha\beta}(x,z) 
&=&-{1\over 2}{\rm Im}\left[G_F^{\mu\nu\alpha\beta}(x,z)\right] \nonumber \\
&=&{1\over 8\pi}\left[u^{\mu\nu\alpha\beta}(x,z)\delta(\sigma(x,z))
-v^{\mu\nu\alpha\beta}(x,z)\theta(-\sigma(x,z))\right], \\
G_{Ret}^{\mu\nu\alpha\beta}(x,z)
&=&2\theta[\Sigma(x),z]\bar G^{\mu\nu\alpha\beta}(x,z), \\
G_{Adv}^{\mu\nu\alpha\beta}(x,z)
&=&2\theta[z,\Sigma(x)]\bar G^{\mu\nu\alpha\beta}(x,z), 
\end{eqnarray}
where 
$\Sigma(x)$ is an arbitrary space-like hypersurface 
containing $x$, and $\theta[\Sigma(x),z]=1
 -\theta[z,\Sigma(x)]$ is equal to $1$ when $z$ lies in the past 
of $\Sigma(x)$ and vanishes when $z$ lies in the future.


\subsection{Metric perturbation}

Using the above obtained retarded Green function, 
we compute the trace-reversed metric perturbation $\psi^{\mu\nu}(x)$ 
induced by a point-like particle. We assume the energy-momentum tensor
of the form,
\begin{eqnarray}
T^{\mu\nu} &=& T^{\mu\nu}_{(mono)} +T^{\mu\nu}_{(spin)} \,,
\label{eq:point}
\\ 
&& T^{\mu\nu}_{(mono)}(x) = m\int dT v^\mu(x,T) v^\nu(x,T) 
{\delta^{(4)}(x-z(T))\over \sqrt{-g}} \,,
\\ 
&& T^{\mu\nu}_{(spin)} = -m \int dT \nabla_\xi\left(S^{\xi(\mu}(x,T)
v^{\nu)}(x,T){\delta^{(4)}(x-z(T))\over \sqrt{-g}}\right) \,, 
\label{eq:spinTmunu}
\\ 
&& \qquad v^\mu(x,T)=\bar g^\mu{}_\alpha(x,z(T))\dot z^\alpha(T) \,, 
\\
&& \qquad S^{\mu\nu}(x,T)=\bar g^\mu{}_\alpha(x,z(T)) 
\bar g^\nu{}_\beta(x,z(T))S^{\alpha\beta}(T) \,,
\end{eqnarray}
where $\dot z^\alpha(T)=d z^\alpha/dT$, $m$ is the mass of the particle
and $S^{\alpha\beta}(T)$ is an anti-symmetric tensor representing the
specific spin of the particle per unit mass. We call it the spin
tensor of the particle and assume that it satisfies the center of mass
condition,
\begin{eqnarray}
S_{\alpha\beta}(T)\dot z^\beta(T)=0 \,.
\end{eqnarray}

In Chapter 1, section 11, we have given the energy-momentum tensor of a
spinning test particle.\footnote{Note that $S_{\alpha\beta}$ there
  corresponds to $m S_{\alpha\beta}$ here.} 
There the four-velocity of the orbit $v^\alpha=\dot z^\alpha$ is
distinguished from the specific four-momentum of the particle
$u^\alpha=p^\alpha/m$. The difference is $O(S^2/L^2)$ where $S$ is the
magnitude of the spin tensor
$S:=\sqrt{S_{\alpha\beta}S^{\alpha\beta}/2}$. Here we ignore this
difference because of the following reason. 
Since the particle is assumed to represent a black hole,
$m$ will be identified with the black hole mass and $S$ 
with the Kerr spin parameter $a$. Therefore $S$ is assumed to be of
order $Gm$, hence the difference between $v^\alpha$ and $u^\alpha$ is
$O((Gm/L)^2)$. Since we are interested in the radiation reaction of
$O(Gm/L^2)$ to the equations of motion, we may consistently neglect this
difference. 

At this point, we must comment on the reason why 
we may assume the point-like particle for the source. 
Even in the linear perturbation, in order 
to generate a general gravitational field 
in the external zone, we need to consider a source 
with arbitrary higher multipole moments.\footnote{A distributional form
of the energy-momentum tensor with arbitrary higher multipole moments 
was discussed by Dixon\cite{Dixon}.}
However, the $\ell$-th moment of the gravitational field will be
$O((Gm/|X|)^{\ell+1})$ in the matching region if the particle represents 
a black hole. As we shall see in the following discussions, we find it
is not necessary to consider the matchings at $O\bigl((Gm)^3\bigr)$ or
higher in order to derive the equations of motion with the reaction
force of $O(Gm/L^2)$. Hence the moments higher than the spin can be
consistently neglected.

We should also note that the metric perturbation induced by
$T^{\mu\nu}_{(spin)}$ is $O\bigl((Gm)^2\bigr)$ for $S=O(Gm)$. 
At first glance, one might think that this implies the necessity of the
second order perturbation theory if we are to incorporate the spin
effect of the particle in the expansion with respect to $Gm$ in a
consistent way. However, provided that the construction of the metric by
the matched asymptotic expansion is consistent, the second order
perturbation theory turns out to be unnecessary. In fact, we shall find
that the spin-induced metric perturbation of $O\bigl((Gm)^2\bigr)$ gives
rise to the leading order spin-curvature coupling term of $O(Gm/L^2)$ in
the equations of motion, while the spin-independent metric perturbation
of $O\bigl((Gm)^2\bigr)$ does not contribute to the reaction force term
at $O(Gm/L^2)$.

Without any further approximation, 
the metric perturbation due to the point-like particle becomes 
\begin{eqnarray}
\psi^{\mu\nu}(x)&=&
2Gm\Biggl(\biggl[ 
{1\over\dot\sigma(x,z(T))}u^{\mu\nu}{}_{\alpha\beta}(x,z(T)) 
\dot z^\alpha(T) \dot z^\beta(T) 
\nonumber \\ && \qquad \quad 
+{\ddot\sigma(x,z(T))\over\dot\sigma^3(x,z(T))} 
u^{\mu\nu}{}_{\alpha\beta}(x,z(T))\sigma_{;\gamma}(x,z(T)) 
S^{\gamma\alpha}(T)\dot z^\beta(T) 
\nonumber \\ && \qquad \quad 
+{1\over\dot\sigma(x,z(T))}u^{\mu\nu}{}_{\alpha\beta;\gamma}(x,z(T)) 
S^{\gamma\alpha}(T) \dot z^\beta(T) 
\nonumber \\ && \qquad \quad 
-{1\over\dot\sigma^2(x,z(T))}{d\over dT}\left(
u^{\mu\nu}{}_{\alpha\beta}(x,z(T))\sigma_{;\gamma}(x,z(T))
S^{\gamma\alpha}(T) \dot z^\beta(T) \right)
\nonumber \\ && \qquad \quad 
+{1\over\dot\sigma(x,z(T))}v^{\mu\nu}{}_{\alpha\beta}(x,z(T))
\sigma_{;\gamma}(x,z(T)) S^{\gamma\alpha}(T)\dot z^\beta(T) 
\biggr]_{T=T_{Ret}(x)} 
\nonumber \\ && \qquad \quad 
-\int^{T_{Ret}(x)}_{-\infty}
dT \biggl(v^{\mu\nu}{}_{\alpha\beta}(x,z(T))
\dot z^\alpha(T)\dot z^\beta(T)
\nonumber \\ && \qquad \qquad \qquad \quad 
+v^{\mu\nu}{}_{\alpha\beta;\gamma}(x,z(T))
S^{\gamma\alpha}(T)\dot z^\beta(T)\biggr)
\Biggr) \,,
\label{eq:metper} 
\end{eqnarray} 
where $T_{Ret}(x)$ is the retarded time 
of the particle and is a scalar function which is determined by 
\begin{eqnarray}
\sigma\left(x,z(T_{Ret})\right)=0 \,, \quad
\theta\left(\Sigma(x),z(T_{Ret})\right)=1 \,.
\end{eqnarray}

Since the retarded time $T_{Ret}(x)$ is not convenient for specifying
the field point $x$ around the particle trajectory in the following
computations, we introduce a new specification of $x$ as follows.
We foliate the spacetime with spacelike 3-surfaces 
perpendicular to the particle trajectory. 
Specifically, the 3-surfaces are defined as a one-parameter 
family of $T$ by the relation,
$\sigma_{;\alpha}(x,z(T))\dot z^\alpha(T)=0$. 
We denote the value of $T$ of the 3-surface containing 
the point $x$ by $T_x$. 
That is
\begin{eqnarray}
\sigma_{;\alpha}(x,z(T_x))\dot z^\alpha(T_x)=0 \,, 
\label{eq:foli}
\end{eqnarray}
where we have introduced the notation, 
\begin{eqnarray}
Q_{;\alpha}(x,z(T_x)) & := & [Q_{;\alpha}(x,z)]_{z=z(T_x)} \,, 
\cr
Q_{;\mu}(x,z(T_x)) & := & [Q_{;\mu}(x,z)]_{z=z(T_x)} \,.
\end{eqnarray}
Note that
\begin{equation}
\left[Q(x,z(T_x))\right]_{;\mu} = Q_{;\mu}(x,z(T_x)) 
  +Q_{;\alpha}(x,z(T_x)) \dot z^{\alpha}(T_x) T_{x;\mu} \,. 
\end{equation}
We use $\sigma_{;\alpha}(x,z(T_x))$ to distinguish the spatial points on
the same 3-surface, and denote the spatial distance from $z(T_x)$ to $x$
by 
\begin{equation}
\epsilon(x):=\sqrt{2\sigma(x,z(T_x))} \,.
\end{equation} 
In the matching region, we have
\begin{eqnarray}
Gm \ll \epsilon(x) \ll L.
\end{eqnarray}

To obtain the external metric in the matching region, we first
consider the $\epsilon$-expansion of the time retardation,
$\delta_{Ret}(x)$,
\begin{eqnarray}
\delta_{Ret}(x):=T_{Ret}(x)-T_x \,. 
\end{eqnarray}
It is given by expanding Eq.~(\ref{eq:foli}) as
\begin{eqnarray}
0 &=& \left[\sigma(x,z(T))\right]_{\tau=T_{Ret}(x)}
\nonumber \\ 
&=&\sigma(x,z(T_x))+\dot\sigma(x,z(T_x))\delta_{Ret}(x)
\nonumber \\ && \qquad 
+{1\over 2}\ddot{\sigma}(x,z(T_x))\delta_{Ret}^2(x)
+{1\over 3!}\stackrel{...}{\sigma}(x,z(T_x))\delta_{Ret}^3(x)
\nonumber \\ && \qquad 
+{1\over 4!}\stackrel{....}{\sigma}(x,z(T_x))\delta_{Ret}^4(x)
+O(\epsilon^5) \,. 
\end{eqnarray}
Using Eqs.~(\ref{128db}), (\ref{eq:foli}), 
and the normalization condition, $(dz/dT)^2 = -1+ O(Gm/L)$, which will
be proved to be consistent later, each term in the above is computed as
\begin{eqnarray}
\sigma(x,z(T_x))&=&{1\over 2}\epsilon^2(x) \,,
\\
\dot\sigma(x,z(T_x))
&=&\sigma_{;\alpha}(x,z(T_x))\dot z^{\alpha}(T_x)=0 \,,
\label{eq:Afoli}\\
\ddot\sigma(x,z(T_x))\,&=:&-\kappa^2(x)
\nonumber\\
&=&\sigma_{;\alpha\beta}(x,z(T_x))\dot z^\alpha(T_x)\dot z^\beta(T_x)
+\sigma_{;\alpha}(x,z(T_x))\ddot z^{\alpha}(T_x)
\nonumber \\ 
& = & 
\left(g_{\alpha\beta}(z(T_x))-{1\over 3}
 R_{\alpha}{}^{\gamma}{}_{\beta}{}^{\delta}(z(T_x)) 
 \sigma_{;\gamma}(x,z(T_x)) \sigma_{;\delta}(x,z(T_x))\right)
  \dot z^{\alpha}(T_x)\dot z^{\beta}(T_x)
\nonumber \\ 
&& \quad +
 \sigma_{;\alpha}(x,z(T_x)) \ddot z^{\alpha}(T_x)
 +O(\epsilon^3) \,,
\\ 
\stackrel{...}{\sigma}(x,z(T_x))
&=&\sigma_{;\alpha}(x,z(T_x))\stackrel{...}{z}{}^\alpha(T_x)
+O(\epsilon^2) \,,
\\ 
\stackrel{....}{\sigma}(x,z(T_x))
&=& -g_{\alpha\beta}(z(T_x))
\ddot z^\alpha(T_x)\ddot z^\beta(T_x)+O(\epsilon) \,. 
\end{eqnarray}
where we have introduced $\kappa(x)$ to denote 
$\sqrt{-\ddot\sigma(x,z(T_x))}$.
{}From these, we obtain 
\begin{eqnarray}
\delta_{Ret}(x)&=&
-\epsilon(x)\kappa^{-1}(x)\biggl(1
-{1\over 6}\epsilon(x)\kappa^{-3}(x)
\stackrel{...}{z}{}^{\alpha}(T_x)\sigma_{;\alpha}(x,z(T_x))
\nonumber \\ && \qquad \qquad
-{1\over 24}\epsilon^2(x)\kappa^{-4}(x)\ddot z^2(T_x)\biggr)
+O(\epsilon^4) \,.
\label{eq:retard}
\end{eqnarray} 

With the help of Eq.~(\ref{eq:retard}), we then obtain the
expansion of various terms in Eq.~(\ref{eq:metper}).
We have
\begin{eqnarray}
&& \left[{1\over\dot\sigma(x,z(T))}\right]_{T=T_{Ret}(x)}
\nonumber \\ && \qquad 
={1\over\epsilon(x)\kappa(x)}\biggl(1
-{1\over 3}\epsilon(x)\stackrel{...}{z}{}^\alpha(T_x)
\sigma_{;\alpha}(x,z(T_x))
-{1\over 8}\epsilon^2(x)\ddot z^2(T_x)
+O(\epsilon^3)\biggr) \,.
\label{eq:sigmadot}
\end{eqnarray} 
In order to obtain the expansion of $u^{\mu\nu\,\alpha\beta}(x,z)$ given 
by Eq.~(\ref{eq:u}), we also need the following expansions:
\begin{eqnarray}
&& \biggl[\Delta^{1/2}(x,z(T))\biggr]_{T=T_{Ret}(x)}
=1+O(\epsilon^3) \,,
\\
&& \biggl[\bar g_{\mu\alpha}(x,z(T))\biggr]_{T=T_{Ret}(x)}
\nonumber \\ && \qquad 
=\bar g_{\mu\alpha}(x,z(T_x))
-{1\over 2}\bar g_\mu{}^\beta(x,z(T_x))
R_{\alpha\beta\gamma\delta}(z(T_x))
\sigma^{;\gamma}(x,z(T_x))\dot z^\delta(T_x)\epsilon(x)
\nonumber \\ && \qquad \qquad
+O(\epsilon^3) \,, 
\\
&& \biggl[\dot z^\alpha(T)\biggr]_{T=T_{Ret}(x)} 
=\dot z^\alpha(T_x)-\epsilon(x)\kappa^{-1}(x)\ddot z^\alpha(T_x) 
+{1\over 2}\epsilon^2(x)\stackrel{...}{z}{}^\alpha(T_x)+O(\epsilon^3) \,. 
\label{eq:expansions}
\end{eqnarray}

In the above expressions there appear higher derivatives of $\dot z$, 
such as $\ddot z$ and $\stackrel{...}{z}$, where 
a dot means the covariant derivative $D/dT$ along the trajectory of 
the particle. Since we are considering the case in which the radiation
reaction force is $O(Gm/L^2)$, it is reasonable to assume these
derivatives are smaller by a factor of $O(1/T_r)$, i.e., 
\begin{eqnarray}
{D^{n+1}z(T)\over dT^{n+1}}\sim {1\over T_r L^{n-1}}
<O\left({\epsilon(x)\over L^{n+1}}\right)\quad(n\ge1),
\end{eqnarray}
where $T_r=O(L^2/(Gm))$ is the reaction time scale.
We shall find that this is consistent with the equations of motion in
the end. 

Keeping this fact in mind, and using Eqs.~(\ref{eq:sigmadot}) $\sim$
(\ref{eq:expansions}), we obtain the $\epsilon$-expansion of the
trace-reversed metric perturbation, Eq.~(\ref{eq:metper}), as
\begin{equation}
  \psi^{\mu\nu}=\psi^{\mu\nu}_{(mono)}+\psi^{\mu\nu}_{(spin)}
+\psi^{\mu\nu}_{(tail)} \,,
\label{eq:metper0} \\ 
\end{equation}
where
\begin{eqnarray}
\psi^{\mu\nu}_{(mono)}(x)
&=& 2Gm \bar g^\mu_\alpha \bar g^\nu_\beta
\Biggl({2\over \epsilon}\kappa^{-1}\dot z^\alpha\dot z^\beta
\nonumber \\ && \quad 
-4\dot z^{(\alpha}\ddot z^{\beta)}
+2\dot z^\gamma\sigma^{;\delta}\dot z^\epsilon 
R_{\gamma\delta\epsilon}{}^{(\alpha}\dot z{}^{\beta)} 
-2\epsilon R^\alpha{}_\gamma{}^\beta{}_\delta\dot z^\gamma\dot z^\delta 
+O(\epsilon^2)\Biggr),
\\
\psi^{\mu\nu}_{(spin)}(x)
&=& -4Gm \bar g^\mu_\alpha \bar g^\nu_\beta
\Biggl({1\over \epsilon^3}\dot z^{(\alpha} S^{\beta)\gamma}\sigma_{;\gamma}
+O((Gm)\epsilon^0)\Biggr), 
\\
\psi^{\mu\nu}_{(tail)}(x)
&=& 2Gm \bar g^\mu_\alpha \bar g^\nu_\beta
\nonumber \\ && \times
\Biggl(-\int^{T_x}_{-\infty}dT'
\biggl(v^{\alpha\beta}{}_{\alpha'\beta'}(z(T_x),z(T'))
\dot z^{\alpha'}(T')\dot z^{\beta'}(T')
\nonumber \\ && \quad 
+v^{\alpha\beta}{}_{\alpha'\beta';\gamma'}(z(T_x),z(T'))
S^{\gamma'\alpha'}(T')\dot z^{\beta'}(T')\biggr)
\nonumber \\ && \qquad
+\sigma_{;\gamma}\int^{T_x}_{-\infty}dT'
\biggl(v^{\alpha\beta}{}_{\alpha'\beta'}{}^{;\gamma}
(z(T_x),z(T'))\dot z^{\alpha'}(T')\dot z^{\beta'}(T')
\nonumber \\ && \qquad \quad 
+v^{\alpha\beta}{}_{\alpha'\beta';\gamma'}{}^{;\gamma}
(z(T_x),z(T'))S^{\gamma'\alpha'}(T')\dot z^{\beta'}(T')\biggr)
+O(\epsilon^2)\Biggr),
\end{eqnarray}
where $\bar g^\mu_\alpha=\bar g^\mu_\alpha(x,z(T_x))$.
The part $\psi_{(tail)}^{\mu\nu}$ is called the tail term because it is
not due to the direct light cone propagation of waves but due to
multiple curvature scattering of waves as described by the
$v^{\mu\nu\alpha\beta}(x,z)$ term in the Green function.


\subsection{Transformation to the internal coordinates}

In order to write down the external metric in terms of the
internal coordinates, 
we consider a coordinate transformation from $x$ to $\{X^a\}$ 
given in the form,
\begin{eqnarray}
\sigma_{;\alpha}(x,z(T))=-{\cal F}_\alpha(T,X) \label{eq:trans}. 
\end{eqnarray}
We restrict our consideration on a coordinate transformation which
satisfies the following requirements. 
We assume $X^i=0$ corresponds to the center of the particle, 
$x^\alpha=z^\alpha(T)$, hence ${\cal F}_\alpha=0$ at $X^i=0$. 
We also assume that the right hand side of Eq.~(\ref{eq:trans}) can 
be expanded in positive powers of $X^i$ as 
\begin{eqnarray}
{\cal F}_\alpha(T,X)=f_{\alpha i}(T)X^i+{1\over 2}f_{\alpha ij}(T)X^iX^j
+{1\over 3!}f_{\alpha ijk}(T)X^iX^jX^k+\cdots.
\label{eq:calFexpand}
\end{eqnarray}
Although it is possible that there appear more complicated terms such as
$X^iX^j/|X|$, we simply ignore such kinds of terms. We shall find it is
consistent within the order of the approximation to which we are going
to develop our consideration. Here $f_{\alpha i_1 \cdots i_n}(T)$ is
totally symmetric for $i_1\cdots i_n$ and is at most of
$O(L^{-(n-1)})$. Using Eqs.~(\ref{128db}) and (\ref{173db}), the total
derivative of Eq.~(\ref{eq:trans}) gives the important relation, 
\begin{eqnarray}
\bar g^\alpha{}_\mu(z(T),x)dx^\mu &=&
\Biggl( {dz^\alpha \over dT}(T) +{Df^\alpha{}_i\over dT}(T)X^i
+{1\over 2}{Df^\alpha{}_{ij}\over dT}(T)X^iX^j
\nonumber \\ && \qquad \quad
-{1\over 2}R^\alpha{}_{\beta\gamma\delta}(z(T))
f^\beta{}_i(T){d z^\gamma\over dT}(T)
f^\delta{}_j(T)X^iX^j\Biggr) dT
\nonumber \\ &&
+\Biggl( f^\alpha{}_i+f^\alpha{}_{ij}(T)X^j
+{1\over 2}f^\alpha{}_{ijk}(T)X^jX^k
\nonumber \\ && \qquad \quad
-{1\over 6}R^\alpha{}_{\beta\gamma\delta}(z(T))
f^\beta{}_j(T)f^\gamma{}_i(T)
f^\delta{}_k(T)X^jX^k\Biggr) dX^i. 
\nonumber \\ && \qquad 
+O(|X|^3)
\label{eq:dtrans}
\end{eqnarray}

In the following sections, we write down the external metric 
in terms of the internal coordinates in the matching region to obtain
the equations of motion. 


\section{Equations of motion for a monopole particle}
\label{sec:deri1}
In this section, we adopt the method (1)
mentioned in section \ref{sec:intro} to derive the equations of
motion. We restrict our consideration to the case of a monopole
particle, which is necessary because we use a well-established method
to decompose the metric in the internal scheme by the tensor harmonics. 
The tensor harmonics are classified by the total 
angular momentum, $J$, reflecting the spherical symmetry 
of the Schwarzschild black hole. 

In the internal scheme, the monopole mode $(J=0)$ corresponds 
to the mass perturbation. Thus we may set this mode to zero
since it is natural to suppose that the change of mass due to the
radiation reaction is negligible. The dipole modes $(J=1)$ are related
to the translation and rotation. 
The translation modes are purely gauge and thus 
we set them to zero to fix the center of the black hole. 
As we are considering a non-rotating black hole, 
we also set the notational modes to zero. 
In general, the higher modes contain
gauge degrees of freedom as well as the physical ones. 
However, for these higher modes, 
we do not give any principle to fix the gauge for the moment.

Before the explicit computation of the $({}^{m}_{n})$ matching condition, 
we briefly review the construction of the scalar 
and vector harmonics in terms of the 
symmetric trace-free (STF) tensor \cite{BlaDam}.


\subsection{Spherical harmonics expansion}
\label{sssec:harmonics}

We introduce the notation, 
\begin{equation}
 A_{<i_1 i_2\cdots i_\ell>}, 
\end{equation}
to represent the totally symmetric trace-free 
part of $A_{i_1 i_2\cdots i_\ell}$.
More explicitly in the cases of $\ell=2$, $3$, 
\begin{eqnarray}
 A_{<ij>} & = & A_{(ij)}-{1\over 3}\delta_{ij} A_{kk}, \cr
 A_{<ijk>} & = & A_{(ijk)}-{1\over 5}\left(
 \delta_{ij} A_{(kmm)}+\delta_{jk} A_{(imm)}+
 \delta_{ki} A_{(jmm)}\right).
\end{eqnarray}

The spherical harmonics expansion of a scalar function 
$A$ on the unit-sphere can be written as 
\begin{equation}
 A=\sum_{\ell=0}^{\infty} A_{<i_1 i_2\cdots i_\ell>}
 n^{<i_1}n^{i_2}\cdots n^{i_\ell>}, 
\end{equation}
where $n^i=X^i/|X|$. 
In this case, the order $\ell$, which is associated with the angular
dependence, is equivalent to the total angular momentum, $J$. 
Thus the $J$ mode of the $(TT)$-component 
of the metric perturbation is totally determined by 
its angular dependence. 
Namely, the terms in the $(TT)$-component 
of the metric perturbation which contain
\begin{equation}
 1,\quad n^i,\quad n^{<i}n^{j>}, 
\end{equation}
correspond to the $J=0$, $1$, $2$ modes, respectively. 

Next we consider the expansion of a vector field $A_i$, 
\begin{equation}
 A_i=\sum_{\ell=0}^{\infty} A_{i <i_1 i_2\cdots i_\ell>}
 n^{<i_1}n^{i_2}\cdots n^{i_\ell>}. 
\end{equation}
In this case the term of the $\ell$-th order in the angular dependence
is decomposed into $J=\ell+1$, $\ell$ and $\ell-1$.
This is done by using the Clebsch-Gordan reduction formula \cite{BlaDam}, 
\begin{equation}
 U_i T_{i_1 i_2\cdots i_\ell}= R^{(+)}_{i<i_1 i_2\cdots i_\ell>}
  +{\ell\over \ell+1} \epsilon_{ji<i_\ell} 
   R^{(0)}_{i_1 i_2\cdots i_{\ell-1}>j}
  +{2\ell-1\over 2\ell+1} \delta_{i<i_\ell} 
   R^{(-)}_{i_1 i_2\cdots i_{\ell-1}>}, 
\end{equation}
where $T_{i_1 i_2\cdots i_\ell}$ is a STF tensor of order $\ell$ and 
\begin{eqnarray}
    R^{(+)}_{i_1 i_2\cdots i_{\ell+1}} & := & 
    U_{<i_{\ell+1}} T_{i_1 i_2\cdots i_\ell>},
\cr
    R^{(0)}_{i_1 i_2\cdots i_{\ell}} & := & 
     U_{j} T_{k<i_1 i_2\cdots i_{\ell-1}} \epsilon_{i_\ell>jk}, 
\cr 
    R^{(-)}_{i_1 i_2\cdots i_{\ell-1}} & := & 
     U_{j} T_{j i_1 i_2\cdots i_{\ell-1}}. 
\end{eqnarray}
We perform the decomposition explicitly for $\ell\le 2$ here. 
For $\ell=0$, there exists no $J=0$ mode and 
it trivially corresponds to the $J=1$ mode. 
For $\ell=1$, the decomposition is performed as 
\begin{equation}
 A_{ij}n^j= \left[\left(A_{(ij)}
  -{1\over 3}\delta_{ij} A_{kk}\right) + A_{[ij]} 
  +{1\over 3}\delta_{ij} A_{kk}\right] n^j, 
\end{equation}
and the first, second and third terms in the square brackets 
correspond to the $J=2$ ,$1$ and $0$ modes, respectively. 
{}For $\ell=2$, we obtain the decomposition formula as 
\begin{equation}
 A_{i<jk>}n^{<j} n^{k>}= \left[A_{<ijk>}+{2\over 3}
  \epsilon_{mi<j} B^{(2)}_{k>m} + {3\over 5} 
  \delta_{i<j} B^{(1)}_{k>}\right]
  n^{<j} n^{k>}, 
\label{c10}
\end{equation}
where
\begin{eqnarray}
 B^{(2)}_{ij} & = & {1\over 2}
 (A_{k<mi>}\epsilon_{jkm}+A_{k<mj>}\epsilon_{ikm}),
\cr 
 B^{(1)}_{k} & = & A_{i<jk>}\delta_{ij}\,,
\label{c9}
\end{eqnarray}
and the first, second and third terms correspond to the $J=3$, $2$
and $1$ modes, respectively.

We omit the general discussion on the expansion of the tensor field 
and we shall give a specific argument when necessary.


\subsection{Geodesics; $({}^0_0)$ and $({}^{1}_0)$ matching}
\label{sssec:geodesics}
We begin with the $({}^0_0)$ and $({}^{1}_0)$ matchings
which are, respectively, of $O((Gm/L)^0)$ and of $O((Gm/L)^{1/2})$ in
the matching region. First we consider the external scheme. 
In these matchings the external metric is the background itself. 
Here, the necessary order of expansion in $|X|$ is $O(|X|)$. 
We note 
\begin{eqnarray}
g_{\mu\nu}(x)dx^\mu dx^\nu =g_{\alpha\beta}(z)\bar g^\alpha{}_\mu(z,x)
\bar g^\beta{}_\nu(z,x)dx^\mu dx^\nu \,.
\end{eqnarray}
Then from Eq.~(\ref{eq:dtrans}), we get 
\begin{eqnarray}
g_{\mu\nu}(x)dx^\mu dx^\nu &=& 
\left( \left({dz \over dT}\right)^2(T) 
+2{dz^\alpha \over dT}(T){D{f}_{\alpha i} \over dT}(T)X^i
\right)dT^2
\nonumber \\ && 
+2\left( {dz^\alpha \over dT}(T) {f}_{\alpha i}(T)
+{dz^\alpha \over dT}(T) {f}_{\alpha ij}(T)X^j
+{f}^{\alpha}{}_{i}(T){D{f}_{\alpha j} \over dT}(T)X^j
\right)dTdX^i
\nonumber \\ && 
+\left({f}^{\alpha}{}_{i}(T){f}_{\alpha j}(T)
+2{f}^{\alpha}{}_{i}(T){f}_{\alpha jk}(T)X^k\right)dX^idX^j
\nonumber \\ && \qquad 
+O\left({|X|^2\over L^2}\right) \,.
\label{eq:ext1}
\end{eqnarray}
Comparing the above equation with Eq.~(\ref{eq:ext}) and
looking at the dependence on $X$,
one can readily extract out ${}^{(0)}_{(0)}h_{ab}$ 
and ${}^{(1)}_{(0)}h_{ab}$ to the lowest order in $Gm/L$.

Next we consider the internal scheme. 
The $({}^0_0)$-component is trivially given 
by the flat Minkowski metric. Hence the $({}^0_0)$ matching becomes
\begin{eqnarray}
-1 &=& \left({dz \over dT}\right)^2(T) 
+O\left({Gm\over L}\right) \,, 
\qquad \mbox{($TT$)-component}, \label{eq:m0tt}
\\ 
0 &=& {dz^\alpha \over dT}(T) f_{\alpha i}(T)
+O\left({Gm\over L}\right) \,, 
\qquad \mbox{($Ti$)-component}, \label{eq:m0ti}
\\ 
\delta_{ij} &=& f^{\alpha}{}_{i}(T)f_{\alpha j}(T)
+O\left({Gm\over L}\right) \,, 
\qquad \mbox{($ij$)-component}. \label{eq:m0ij}
\end{eqnarray}
Equations~(\ref{eq:m0ti}) and (\ref{eq:m0ij}) indicate
that $f^{\alpha i}(T)$ are spatial triad basis 
along the orbit, i.e., 
\begin{equation}
f^\alpha{}_k(T) f^\beta{}_k(T)=g^{\alpha\beta}(z(T))
+{dz^\alpha\over dT}(T){dz^\beta\over dT}(T) 
+O\left({Gm\over L}\right) \,.
\label{eq:triad}
\end{equation}

To know the $({}^1_0)$-component of the internal scheme, 
it is better to consider all the $({}^1_n)$-components at the same time.
Namely we consider the linear perturbation of the black hole
${}^{(1)}H_{ab}$. For this purpose, we consider the harmonic
decomposition of linear perturbation as discussed in subsection 
\ref{sssec:harmonics}. 
Since the time scale associated with the perturbation should be of the
order of the background curvature scale $L$, it is much larger than the
matching radius $(GmL)^{1/2}$. Therefore the perturbation may be
regarded as static. It is known that all the physical static
perturbations regular on the black hole horizon behave as $\sim |X|^{J}$
asymptotically where $J$ is the angular momentum eigenvalue. However, in
${}^{(1)}_{(n)}H_{ab}$, there exists no term which behaves as $\sim
|X|^{m}$ $(m\ge 2)$. Hence, except for gauge degrees of freedom,
${}^{(1)}_{(n)}H_{ab}$ contain only $J=0$, $1$ modes. As mentioned
before, we set the perturbation of $J=0$, $1$ modes to zero. Thus we
conclude that we may set
\begin{eqnarray}
{}^{(1)}_{(n)}H_{ab}=0 \,,
\end{eqnarray}
for all $n$. This is the gauge condition we adopt for the internal
scheme at $O(1/L)$. In particular this condition gives the $({}^1_0)$
matching as
\begin{eqnarray}
0 &=& 
2{dz^\alpha \over dT}(T){D{f}_{\alpha i} \over dT}(T)X^i 
+O\left({Gm\over L}{|X|\over L}\right) \,, 
\qquad \mbox{($TT$)-component}, \label{eq:m1/2tt}
\\
0 &=& 
{dz^\alpha \over dT}(T) {f}_{\alpha ij}(T)X^j
+{f}^{\alpha}{}_{i}(T){D{f}_{\alpha j} \over dT}(T)X^j 
\nonumber \\ && \qquad 
+O\left({Gm\over L}{|X|\over L}\right) \,, 
\qquad\qquad\qquad\qquad\quad \mbox{($Ti$)-component}, \label{eq:m1/2ti}
\\ 
0 &=& 
2{f}_{\alpha (i}(T){f}^{\alpha}{}_{j)k}(T)X^k 
+O\left({Gm\over L}{|X|\over L}\right) \,, 
\qquad \mbox{($ij$)-component}. \label{eq:m1/2ij}
\end{eqnarray}

Then the covariant $T$-derivative 
of Eq.~(\ref{eq:m0tt}) and that of Eq.~(\ref{eq:m0ti}) 
with Eq.~(\ref{eq:m1/2tt}) 
result in the background geodetic motion, 
\begin{eqnarray}
{D\over dT}\left({dz^\alpha \over dT}\right)(T)=
O\left({Gm\over L}{1\over L}\right) \,.
\label{eq:geo}
\end{eqnarray}
One can see from Eq.~(\ref{eq:m0tt}) that the internal time coordinate
$T$ becomes a proper time of the orbit in the lowest order in $Gm/L$.  
In the same manner, Eq.~(\ref{eq:m1/2tt}) and 
the covariant $T$-derivative of Eq.~(\ref{eq:m0ij}) 
with $(ij)$-antisymmetric part of Eq.~(\ref{eq:m1/2ti}) give 
the geodetic parallel transport of the triad ${f}^{\alpha}{}_{i}(T)$, 
\begin{eqnarray}
{D\over dT}{f}^{\alpha}{}_{i}(T)
=O\left({Gm\over L}{1\over L}\right) \,. 
\label{eq:para}
\end{eqnarray}
Further, from Eqs.~(\ref{eq:m1/2ti}) and (\ref{eq:m1/2ij}), 
we can see  
\begin{eqnarray}
{f}^{\alpha}{}_{ij}(T)
=O\left({Gm\over L}{1\over L}\right) \,. 
\label{eq:faij}
\end{eqnarray}


\subsection{Hypersurface condition; $({}^2_0)$ matching}
\label{sssec:hypersurface}
We now proceed to the $({}^{2}_{0})$ matching, in which the external
metric is still given by the background but there appear
non-trivial perturbations in the internal scheme.
Although it is of $O(Gm/L)$ in the matching region and 
$O((Gm/L)^{1/2})$ higher than the remaining $({}^0_1)$-component,
we consider it first for the reason which will be clarified below.

In order to obtain $\displaystyle {}^{(2)}_{(0)}h_{ab}$,
we expand the external metric in terms of the internal coordinates up to
$O(|X|^2)$, i.e., we have to go one order higher than
Eq.~(\ref{eq:ext1}). Then the $(^2_0)$ matching becomes
\begin{eqnarray}
{1\over L^2}{}^{(2)}_{(0)}H_{TT}
&=& -R_{\alpha\beta\gamma\delta}(z(T)){dz^\alpha\over dT}(T)
f^\beta{}_i(T){dz^\gamma\over dT}(T)f^\delta{}_j(T)X^i X^j
\nonumber \\
 && +O\left({Gm\over L}{|X|^2\over L^2} \right),
\qquad \mbox{($TT$)-component},
\label{eq:h20TT}\\ 
{1\over L^2}{}^{(2)}_{(0)}H_{Ti} 
&=& {1\over 2}{dz^\alpha \over dT}(T) {f}_{\alpha ijk}(T) X^j X^k
\nonumber\\
&& -{2\over 3}R_{\alpha\beta\gamma\delta}(z(T))
{dz^\alpha\over dT}(T){f}^{\beta}{}_{j}(T)
{f}^{\gamma}{}_{i}(T){f}^{\delta}{}_{k}(T)X^j X^k
\nonumber\\
&&+O\left({Gm\over L}{|X|^2\over L^2} \right),
\qquad \mbox{($Ti$)-component},
\label{eq:h20Ti} \\ 
{1\over L^2}{}^{(2)}_{(0)}H_{ij} 
&=& {f}_{\alpha (i}(T){f}^{\alpha}{}_{j)kl}(T)X^k X^l 
\nonumber\\
&&-{1\over 3}R_{\alpha\beta\gamma\delta}(z(T))
{f}^{\alpha}{}_{i}(T){f}^{\beta}{}_{k}(T)
{f}^{\gamma}{}_{j}(T){f}^{\delta}{}_{l}(T)X^k X^l
\nonumber \\ 
&&+O\left({Gm\over L}{|X|^2\over L^2} \right),
\qquad \mbox{($ij$)-component},
\label{eq:h20ij}
\end{eqnarray}
where Eqs.~(\ref{eq:para}) and (\ref{eq:faij}) have been used to
simplify the expressions. Since we have set ${}^{(1)}_{(n)}H_{ab}=0$,
the first non-trivial perturbations of the internal metric appear in
${}^{(2)}_{(n)}H_{ab}$. Hence they describe the linear perturbation of
the black hole metric in the internal scheme. Then we have to fix the
gauge condition for this perturbation to perform the matching.
{}For ${}^{(2)}_{(0)}H_{ab}$, since the physical perturbation contained
in it is quadrupolar, we fix the gauge so that all the $J$ modes except
$J=2$ are zero. Then the $({}^2_0)$ matching becomes as follows.

First consider the $(TT)$-component of the metric.
The right hand side of Eq.~(\ref{eq:h20TT}) may contain $J=0$, $2$
modes. The $J=0$ mode, however, vanishes because of the background Ricci 
flatness. Hence this matching just determines the physical perturbation
in the $(TT)$-component.

As for the $(Ti)$-component, the right hand side of Eq.~(\ref{eq:h20Ti}) 
may contain $J=1$, $2$, $3$ modes. As before, the $J=2$ mode just
determines the physical perturbation of the $(Ti)$-component.
So we put $J=0$, $3$ modes to zero. However, they are found to be absent
in the second term of Eq.~(\ref{eq:h20Ti}). To see this we first
decompose its angular dependence, 
\begin{equation}
 {dz^{\alpha}\over dT} R_{\alpha\beta\gamma\delta}
 f^{\gamma}{}_{i} 
 \left( f^{\beta}{}_{<j} f^{\delta}{}_{k>} X^{<j} X^{k>} 
 +{1\over 3} f^{\beta}{}_{k} f^{\delta}{}_{k} |X|^{2}\right).
\label{c12}
\end{equation}
Using Eq.~(\ref{eq:triad}) and the fact that the Ricci tensor vanishes,
the second term in the parentheses is rewritten as 
\begin{equation}
 {1\over 3} {dz^{\alpha}\over dT} R_{\alpha\beta\gamma\delta}
 f^{\gamma}{}_{i} 
 {dz^{\beta}\over dT} {dz^{\delta}\over dT} |X|^{2}, 
\end{equation}
and is found to be zero due to the symmetry of the Riemann tensor.
The first term in the parentheses of Eq.~(\ref{c12}) is decomposed
further with the aid of the formulas (\ref{c9}) and (\ref{c10}) as
\begin{equation}
 {dz^{\alpha}\over dT} R_{\alpha\beta\gamma\delta}
 \left(f^{\gamma}{}_{<i} 
 f^{\beta}{}_{j} f^{\delta}{}_{k>} +
 {2\over 3}\epsilon_{mi<j} 
 F^{(2)\gamma\beta\delta}_{k>m} +
 {3\over 5} \delta_{i<j} F^{(1)\gamma\beta\delta}_{k>}
 \right)
 X^{<j} X^{k>},
\label{eq:mino103}
\end{equation}
where
\begin{eqnarray}
F^{(2)\gamma\beta\delta}_{ij} & := & 
 {1\over 2}\left( f^{\gamma}{}_{m} f^{\beta}{}_{<n} 
  f^{\delta}{}_{i>} \epsilon_{jmn}
+ f^{\gamma}{}_{m} f^{\beta}{}_{<n} 
  f^{\delta}{}_{j>} \epsilon_{imn}\right), 
\cr
F^{(1)\gamma\beta\delta}_{i} & := & 
 {1\over 2}\left( f^{\gamma}{}_{k} f^{\beta}{}_{i} 
  f^{\delta}{}_{k} + f^{\gamma}{}_{k} f^{\beta}{}_{k} 
  f^{\delta}{}_{i} \right) -{1\over 3} 
  f^{\gamma}{}_{i} f^{\beta}{}_{k} f^{\delta}{}_{k}\,.
\end{eqnarray}
It is easy to see that the first and third terms in the parentheses of
Eq.~(\ref{eq:mino103}) vanish due to the symmetry of the Riemann tensor
and the Ricci flatness. Thus only the $J=2$ mode remains in the second
term in the right hand side of Eq.~(\ref{eq:h20Ti}).

Decomposing the first term in the right hand side of
Eq.~(\ref{eq:h20Ti}) in a similar manner, we find it contains
$J=1$, $3$ modes as well as $J=2$ mode. Putting the $J=1$ mode to zero,
we obtain
\begin{equation}
{dz^\alpha\over dT}(T)f_{\alpha ikk}(T)
=O\left({Gm\over L}{1\over L^2} \right) \,.
\label{eq:fikk}
\end{equation}
Putting the $J=3$ mode to zero gives
\begin{equation}
{1\over2}{dz^\alpha\over dT}(T)f_{\alpha <ijk>}(T)X^jX^k
=O\left({Gm\over L}{|X|^2 \over L^2} \right) \,. 
\end{equation}
Then combining this with Eq.~(\ref{eq:fikk}), we find
\begin{equation}
{dz^\alpha\over dT}(T)f_{\alpha ijk}(T)
=O\left({Gm\over L}{1\over L^2} \right).
\label{eq:fijk}
\end{equation}

{}From Eqs.~(\ref{eq:m0ti}), (\ref{eq:faij}) and (\ref{eq:fijk}),
we find
\begin{equation}
 {dz^\alpha \over dT}(T)\sigma_{;\alpha}\left(x(T,X),z(T)\right)
 =-{dz^\alpha \over dT}(T){\cal F}_{\alpha}(T,X)=
 O\left({|X|^4\over L^4} L\right),
\end{equation}
to the lowest order in $Gm/L$.
Comparing this with the hypersurface condition of $T_x$,
Eq.~(\ref{eq:foli}), one finds that the $T={\rm constant}$ hypersurface
differs from the $T_x={\rm constant}$ hypersurface only
by $O(\epsilon^4)=O(|X|^4)$. It then follows that all the calculations
done in section \ref{sec:minoExt} remain valid even if we replace
Eq.~(\ref{eq:foli}) with
\begin{equation}
\sigma_{;\alpha}(x,z(T_x))\dot z^\alpha(T_x)=O(\epsilon^4/L^3).
\label{eq:hypcond}
\end{equation}
Thus $T$ can be identified $T_x$ to the lowest order in $Gm/L$.
The reason why we have done the $({}^2_0)$ matching prior to the
remaining $({}^0_1)$ matching is to establish this equivalence of 
$T$ and $T_x$.

Turning to the $(ij)$-component, it may contain $J=0\sim4$ modes.
we first note that the second term of Eq.~(\ref{eq:h20ij})
contains only $J=2$ mode. This can be seen as follows.
First, we define the spatial triad components of the Riemann tensor by 
\begin{equation}
 R_{ijkm}:=R_{\alpha\beta\gamma\delta}
 f^{\alpha}{}_{i} f^{\beta}{}_{j} 
 f^{\gamma}{}_{k} f^{\delta}{}_{m}\,.
\end{equation}
Introducing a symmetric tensor defined by 
\begin{equation}
 {\cal R}_{ij}={1\over 4} \epsilon_{ikm} \epsilon_{jns} 
      R_{kmns}, 
\end{equation} 
we can express $R^{ikjm}$ in terms of ${\cal R}_{ij}$ as
\begin{equation}
R_{ijkm} = \epsilon^{nij} \epsilon^{skm} {\cal R}_{ns}\,.
\end{equation}
Then the symmetric tensor ${\cal R}_{ij}$ is decomposed into STF tensors
as 
\begin{equation}
{\cal R}_{ij}={\cal R}_{<ij>}+ 
              {1\over 3}\delta_{ij}{\cal R}_{kk}. 
\label{calRab}
\end{equation}
Counting the number of indices, we find that 
the first and second terms in Eq.~(\ref{calRab}) 
correspond to $J=2$ and $0$ modes, respectively. 
However, again owing to the symmetry of the Riemann tensor and 
the Ricci flatness, the $J=0$ mode vanishes 
and only the $J=2$ mode remains. 
Therefore the gauge condition for the $(ij)$-component implies
\begin{equation}
\left[f_{\alpha(i}(T)f^{\alpha}{}_{j)kl}(T)\right]_{J\ne2}= 
O\left({Gm\over L}{1\over L^2} \right) \,,
\label{eq:fifjkl}
\end{equation}
where $[\cdots]_{J\ne2}$ means the $J\ne2$ parts of the quantity.
This will be used in the $({}^2_1)$ matching below.


\subsection{External perturbation; $({}^{0}_{1})$ matching}

Now we proceed to the first non-trivial order in $Gm/|X|$. 
For this purpose, we must develop the external scheme. 
However, since the time slicing by the 
internal time coordinate $T$ is now identical to that by $T_x$ 
in the lowest order in $Gm/L$, 
we can use the previously obtained formula (\ref{eq:metper0})
for the external metric perturbation.

Among the matchings which becomes of $O((Gm/L)^{1/2})$ in the matching
region, there remains the $({}^{0}_{1})$ matching.
This matching relates the masses of the particle in both schemes. 
Since this matching is independent of $L$, we may regard the background
external metric as if it were flat. As is well-known, the
linear perturbation induced by a point-like particle of mass $m$ in the
flat background spacetime is exactly equal to the asymptotic metric of a
Schwarzschild black hole of mass $m$ in the linear order in $m$. 
This fact indicates that the matching gives a consistency condition
at this order.

In order to directly check the consistency, we rewrite
Eq.~(\ref{eq:metper0}) in terms of the internal coordinates.
Since ${}^{(0)}_{(1)}h_{ab}\sim|X|^{-1}$, we have only to consider the
$\psi_{(mono)}^{\mu\nu}$ term of Eq.~(\ref{eq:metper0}). Using
Eqs.~(\ref{eq:465}), (\ref{eq:dtrans}) and the fact that 
$\epsilon=\sqrt{{\cal F}_\alpha(T,X){\cal F}^\alpha(T,X)}$, we find
\begin{eqnarray}
Gm{}^{(0)}_{(1)}h_{ab}(X)dX^a dX^b 
= Gm\left({2\over |X|}dT^2+{2\over |X|}dX^idX^i\right), 
\label{eq:10match}
\end{eqnarray}
which corresponds to the asymptotic form of the 
Schwarzschild black hole of mass $m$ in the harmonic coordinates. 


\subsection{Radiation reaction; 
$({}^1_1)$ and $({}^2_1)$ matchings}

There are many components which become of 
$O(Gm/L)$ and $O((Gm/L)^{3/2})$ in the matching region.
However, we are interested in the leading order correction 
to the equations of motion with respect to $Gm/L$ and 
we found in the $({}^0_0)$ and $({}^1_0)$ matchings 
that in the lowest order the terms which behave as $\sim |X|^0$ 
or $|X|^1$ determines the motion of the particle. 
Therefore we consider the $({}^1_1)$ and $({}^2_1)$ matchings here. 

In order to perform the $({}^1_1)$ and $({}^2_1)$ matchings,
the calculation we have done to obtain Eq.~(\ref{eq:10match}) must be
extended to the linear order in $|X|$.
Then the $({}^1_1)$ matching equations are found as
\begin{eqnarray}
{Gm\over L}{}^{(1)}_{(1)}H_{TT}
&=& \left\{ \left({dz \over dT}\right)^2(T)+1 \right\}
+ Gm{dz^\alpha\over dT}(T){dz^\beta\over dT}(T){\Theta}_{\alpha\beta}(T)
\nonumber \\ &&
+O\left(\left({Gm\over L}\right)^2\right) \,, 
\qquad \mbox{($TT$)-component}, 
\label{eq:m1tt}\\
{Gm\over L}{}^{(1)}_{(1)}H_{Ti} 
&=& {dz^\alpha \over dT}(T) {f}_{\alpha i}(T)
+ Gm{dz^\alpha\over dT}(T){f}^{\beta}{}_{i}(T){\Theta}_{\alpha\beta}(T)
\nonumber \\ &&
+O\left(\left({Gm\over L}\right)^2\right) \,,
\qquad \mbox{($Ti$)-component},
 \label{eq:m1ti}\\
{Gm\over L}{}^{(1)}_{(1)}H_{ij} 
&=& \left\{{f}^{\alpha}{}_{i}(T){f}_{\alpha j}(T) -\delta_{ij}\right\}
+Gm{f}^{\alpha}{}_{i}(T){f}^{\beta}{}_{j}(T){\Theta}_{\alpha\beta}(T)
\nonumber \\ &&
+O\left(\left({Gm\over L}\right)^2\right) \,, 
\qquad \mbox{($ij$)-component}, 
\label{eq:m1ij}
\end{eqnarray}
and the $({}^2_1)$ matching as
\begin{eqnarray}
{Gm\over L^2}{}^{(2)}_{(1)}H_{TT} 
&=& 2{dz^\alpha \over dT}(T){D{f}_{\alpha i} \over dT}(T)X^i 
\nonumber \\ && 
+Gm\Biggl\{
{dz^\alpha\over dT}(T){dz^\beta\over dT}(T)
f^{\gamma}{}_{i}(T){\Theta}_{\alpha\beta\gamma}(T)X^i
\nonumber \\ && \quad
-{1\over 3|X|^3}f_{\alpha i}(T)f^{\alpha}{}_{jkl}(T)X^i X^j X^k X^l 
\nonumber \\ && \quad
+{5\over 3|X|}R_{\alpha\beta\gamma\delta}(z(T))
{dz^\alpha\over dT}(T){f}^{\beta}{}_{i}(T)
{dz^\gamma\over dT}(T){f}^{\delta}{}_{j}(T)X^i X^j\Biggr\}
\nonumber \\ && \quad
+O\left(\left({Gm\over L}\right)^2{|X|\over L}\right) \,, 
\qquad \mbox{($TT$)-component}, 
\label{eq:m3/2tt} \\
{Gm\over L^2}{}^{(2)}_{(1)}H_{Ti}
&=& {dz^\alpha \over dT}(T) {f}_{\alpha ij}(T)X^j
+{f}^{\alpha}{}_{i}(T){D{f}_{\alpha j} \over dT}(T)X^j
\nonumber \\ && 
+Gm\Biggl\{
{dz^\alpha\over dT}(T){f}^{\beta}_{i}(T)
{f}^{\gamma}_{j}(T){\Theta}_{\alpha\beta\gamma}(T)X^j
\nonumber \\ && \quad 
+2R_{\alpha\beta\gamma\delta}(z(T)){dz^\alpha\over dT}(T)
{f}^{\beta}{}_{i}(T){dz^\gamma\over dT}(T){f}^{\delta}{}_{j}(T)X^j 
\nonumber \\ && \quad 
+{2\over 3|X|}R_{\alpha\beta\gamma\delta}(z(T))
{dz^\alpha\over dT}(T){f}^{\beta}{}_{j}(T) 
{f}^{\gamma}{}_{i}(T) {f}^{\delta}{}_{k}(T)X^j X^k
\Biggr\}
\nonumber \\ && \quad 
+O\left(\left({Gm\over L}\right)^2{|X|\over L}\right) \,, 
\qquad \mbox{($Ti$)-component}, 
\label{eq:m3/2ti}
\end{eqnarray}
where
\begin{eqnarray}
 Gm {\Theta}_{\alpha\beta}(T) & 
:= & h_{(tail)\alpha\beta}(z(T)), 
\cr
 Gm {\Theta}_{\alpha\beta\gamma}(T)& 
:= & h_{(tail)\alpha\beta;\gamma}(z(T)),
\end{eqnarray}
with
\begin{equation}
 h_{(tail)\mu\nu}(x):=\psi_{(tail)\mu\nu}(x) 
   -{1\over 2} g_{\mu\nu}(x)\psi_{(tail)}(x). 
\label{eq:hv}
\end{equation}
Note that $h_{(tail)\mu\nu}(x)$ is the metric 
perturbation due to $v_{\mu\nu\alpha\beta}(x,z)$ in the Green function. 
The ($ij$)-component of the $({}^2_{1})$ matching is not presented here 
since it will not be used in the following discussion. 

As we have discussed in subsection \ref{sssec:geodesics},
we require ${}^{(1)}_{(1)}H_{ab}=0$. 
Thus the right hand sides of Eqs.~(\ref{eq:m1tt}), 
(\ref{eq:m1ti}) and (\ref{eq:m1ij}) must vanish. 
As for ${}^{(2)}_{(1)}H_{ab}$, following the discussion in
subsection \ref{sssec:hypersurface}, we set all the modes except
$J=2$ to zero. 
Inspection of the right hand sides of Eqs.~(\ref{eq:m3/2tt}) and
(\ref{eq:m3/2ti}) reveals that the terms involving the Riemann tensor 
are in the same forms as those appeared in Eqs.~(\ref{eq:h20TT}),
(\ref{eq:h20Ti}) and (\ref{eq:h20ij}). Hence they contain only $J=2$
modes and do not give any matching condition.
Furthermore, from Eq.~(\ref{eq:fifjkl}), all the modes except $J=2$
contained in the term involving $f^{\alpha}{}_{jkl}$ in
Eq.~(\ref{eq:m3/2tt}) vanish at the lowest order in $Gm/L$.
Hence we only have to consider the remaining terms in
Eqs.~(\ref{eq:m3/2tt}) and (\ref{eq:m3/2ti}). The $J=1$ modes are
extracted out to give
\begin{eqnarray} 
0 &=& 
2{dz^\alpha \over dT}(T){D{f}_{\alpha i} \over dT}(T) 
+ Gm\,{dz^\alpha\over dT}(T){dz^\beta\over dT}(T)
{f}^{\gamma}{}_{i}(T){\Theta}_{\alpha\beta\gamma}(T)
\nonumber \\ && 
+O\left(\left({Gm\over L}\right)^2{1\over L}\right),
 \qquad \mbox{($TT$)-component}, 
\label{eq:m3/2ttg}\\
0 &=& 
{f}_{\alpha [i}(T){D{f}^{\alpha}{}_{j]} \over dT}(T)
+ Gm\,{\Theta}_{\alpha\beta\gamma}(T){dz^\alpha\over dT}(T)
 {f}^{\beta}{}_{[i}(T){f}^{\gamma}{}_{j]}(T)
\nonumber \\ &&
+O\left(\left({Gm\over L}\right)^2{1\over L}\right),
\qquad \mbox{($Ti$)-component}.
\label{eq:m3/2tig}
\end{eqnarray}
The $J=0$ mode is absent in the ($TT$)-component, while that in the
($Ti$)-component exists but it just gives 
the equation which determines $(dz^{\alpha}/dT)f_{\alpha ii}$
to the first order in $Gm/L$. 

Taking the covariant $T$-derivative of Eqs.~(\ref{eq:m1tt}) and 
(\ref{eq:m1ti}) and using Eq.~(\ref{eq:m3/2ttg}), 
we obtain the equations of motion with the $O(Gm/L^2)$ correction 
due to the radiation reaction,
\begin{eqnarray}
{D\over dT}{dz^\alpha\over dT}(T) &=& 
-{Gm\over 2}
\left({\Theta}^\alpha{}_{\beta\gamma}(T)
+{\Theta}^\alpha{}_{\gamma\beta}(T)-{\Theta}_{\beta\gamma}{}^\alpha(T)\right)
{dz^\beta\over dT}(T){dz^\gamma\over dT}(T)
\nonumber \\ && \qquad
+O\left(\left({Gm\over L}\right)^2{1\over L}\right) \,. 
\label{eq:damp1}
\end{eqnarray}
Similarly the $O(Gm/L^2)$ correction to the evolution equations of the
`triad' basis, $f^{\alpha}{}_{i}(T)$, are obtained from
the covariant $T$-derivative of Eq.~(\ref{eq:m1ij}), and
Eqs.~(\ref{eq:m3/2ttg}) and (\ref{eq:m3/2tig}). The result is
\begin{eqnarray}
{D\over dT}{f}^{\alpha}{}_{i}(T) &=&
- {Gm\over 2}
\left({\Theta}^\alpha{}_{\beta\gamma}(T)
+{\Theta}^\alpha{}_{\gamma\beta}(T)-{\Theta}_{\beta\gamma}{}^\alpha(T)\right)
{f}^{\beta}{}_{i}(T){dz^\gamma\over dT}(T)
\nonumber \\ && \qquad 
+O\left(\left({Gm\over L}\right)^2{1\over L}\right) \,.
\label{eq:damp2}
\end{eqnarray}

Since the internal time coordinate $T$ is not properly normalized 
in the external metric, we define the proper time, $\tau=\tau(T)$, such
that $(dz/d\tau)^2 =-1$.  It is easy to see that we should choose 
\begin{equation}
 {d\tau\over dT}=1+{Gm\over 2}\Theta_{\alpha\beta}(T)
 {dz^{\alpha}\over d\tau}(T){dz^{\beta}\over d\tau}(T)
 +O\left(\left({Gm\over L}\right)^2\right) \,.
\end{equation}
Since the second term on the right hand side of this equation 
is proportional to the small perturbation induced by the particle, it is
guaranteed to stay small even after a long time interval compared with
the reaction time scale $T_r=O\left((Gm/L)^{-1}L\right)$. 
Then Eq.~(\ref{eq:damp1}) becomes 
\begin{eqnarray}
&&{D\over d\tau}{dz^\alpha\over d\tau}(\tau) 
\nonumber \\ && \qquad
= -{Gm\over 2}\left({dz^\alpha\over d\tau}{dz^\beta\over d\tau}
{dz^\gamma\over d\tau}{dz^\delta\over d\tau}
+2g^{\alpha\beta}(z){dz^\gamma\over d\tau}{dz^\delta\over d\tau}
-g^{\alpha\delta}(z){dz^\beta\over d\tau}{dz^\gamma\over d\tau}
\right)(\tau)~{\Theta}_{\beta\gamma\delta }(\tau)
\nonumber \\ && \qquad \qquad
+O\left(\left({Gm\over L}\right)^2{1\over L}\right) \,.
\label{eq:damp1r}
\end{eqnarray}
Also, the triad basis are not properly normalized 
in the external metric. Thus we define $e^{\alpha}{}_{i}(\tau)$ as
\begin{eqnarray}
e_{\alpha i}(\tau)e^{\alpha}{}_{j}(\tau) &=& \delta_{ij} \,,
\label{eq:normtriad}\\
e^{\alpha}{}_{i}(\tau) &=& (\delta_{ij}+s_{ij})f^{\alpha}{}_{j}
-Gm({dz^{\alpha} / dT})({dz^{\beta} / dT})f^\gamma{}_i
{\Theta}_{\beta\gamma} \,,
\end{eqnarray}
where $s_{ij}$ is of $O(Gm/L)$ and recalling Eq.~(\ref{eq:m1ti}) the
last term is added so as to satisfy the orthonormal condition,
\begin{eqnarray}
e_{\alpha i}(\tau)({dz^{\alpha} / d\tau})(\tau)=0 \,.
\end{eqnarray}
{}From Eq.~(\ref{eq:normtriad}) we find 
\begin{equation}
 s_{ij}=- {Gm\over 2}\Theta_{\alpha\beta}(\tau)
 {f^{\alpha}{}_i}(\tau){f^{\beta}{}_j}(\tau)
 +O\left(\left({Gm\over L}\right)^2\right) \,.
\end{equation}
Again the correction terms in $e^\alpha_i$ are guaranteed to stay small.
Then the evolution equations of the normalized triad
$e^{\alpha}{}_{i}(\tau)$ become 
\begin{eqnarray}
&&{D \over d\tau}e^{\alpha}{}_{i}(\tau) 
\nonumber \\ && \qquad
= - {Gm\over 2} \left({dz^\alpha\over d\tau}{dz^\beta\over d\tau}
e^{\gamma}{}_{i}{dz^\delta\over d\tau} 
+g^{\alpha\beta}(z){dz^\gamma\over d\tau}e^{\delta}{}_{i} 
-g^{\alpha\delta}(z) e^{\beta}{}_{i}{dz^\gamma\over d\tau} 
\right) (\tau)~ {\Theta}_{\beta\gamma\delta }(\tau) 
\nonumber \\ && \qquad \qquad 
+O\left(\left({Gm\over L}\right)^2{1\over L}\right) \,.
\label{eq:damp2r}
\end{eqnarray}


\section{Equations of motion for a spinning particle}
\label{sec:deri2}

In this section, we consider the equations of motion for a spinning
particle. Different from the Schwarzschild case, we cannot make use of
the mode decomposition by the spherical harmonics since the background
in the internal scheme does not have the spherical symmetry. Therefore,
it is quite unclear for us how to fix the gauge in the internal scheme,
and hence we cannot derive the equations of motion by the consistency
condition of matching.

Instead, we here apply the laws of motion and precession 
discussed by Thorne and Hartle\cite{Thorne1}.
As noted in section \ref{sec:intro}, assuming the consistency
between the internal and external schemes, we can make use of the
matching condition to obtain the internal metric from the knowledge of
the external metric. The problem to derive the equations of motion for a
spinning particle was discussed by Thorne and Hartle\cite{Thorne1} and
the spin-induced force was derived. The discussion given below is an
extension of Ref.~\citen{Thorne1} in the sense that we take into account
the effect of radiation reaction to the motion. Both derivations of
the radiation reaction and the spin-induced force are discussed in a
unified manner. 


\subsection{Laws of motion and precession}

The laws of motion and precession\cite{Thorne1} are derived from
the integral identities given in terms of the Landau-Lifshitz
pseudo-tensor, $t^{\alpha\beta}_{L-L}$, and the Landau-Lifshitz
super-potential, $H^{\alpha\mu\beta\nu}_{L-L}$. 
The Einstein equations can be put into the form,
\begin{eqnarray}
H^{\alpha\gamma\beta\delta}_{L-L}{}_{,\gamma\delta}
=16\pi G(-g)\left(T^{\alpha\beta}+t^{\alpha\beta}_{L-L}\right) \,,
\label{eq:minoEieq}
\end{eqnarray}
where
\begin{eqnarray}
H^{\alpha\gamma\beta\delta}_{L-L}
&=&{\bf g}^{\alpha\beta}{\bf g}^{\gamma\delta}
-{\bf g}^{\alpha\delta}{\bf g}^{\gamma\beta} \,,
\\
(-g)t^{\alpha\beta}_{L-L}
&=&{1\over 16\pi}\biggl\{
{\bf g}^{\alpha\beta}{}_{,\gamma}{\bf g}^{\gamma\delta}{}_{,\delta}
-{\bf g}^{\alpha\gamma}{}_{,\gamma}{\bf g}^{\beta\delta}{}_{,\delta}
+{1\over 2}g^{\alpha\beta}g_{\gamma\delta}
{\bf g}^{\gamma\epsilon}{}_{,\zeta}{\bf g}^{\delta\zeta}{}_{,\epsilon}
\nonumber \\ && \quad
-\left(g^{\alpha\gamma}g_{\delta\epsilon}
{\bf g}^{\beta\delta}{}_{,\zeta}{\bf g}^{\epsilon\zeta}{}_{,\gamma}
+g^{\beta\gamma}g_{\delta\epsilon}
{\bf g}^{\alpha\delta}{}_{,\zeta}{\bf g}^{\epsilon\zeta}{}_{,\gamma}
\right)
+g_{\gamma\delta}g^{\epsilon\zeta}
{\bf g}^{\alpha\gamma}{}_{,\epsilon}{\bf g}^{\beta\delta}{}_{,\zeta}
\nonumber \\ && \quad
+{1\over 8}\left(2g^{\alpha\gamma}g^{\beta\delta}
-g^{\alpha\beta}g^{\gamma\delta}\right)
\left(2g_{\epsilon\zeta}g^{\eta\theta}
-g^{\epsilon\eta}g^{\zeta\theta}\right)
{\bf g}^{\epsilon\eta}{}_{,\gamma}{\bf g}^{\zeta\theta}{}_{,\delta}
\biggr\} \,,
\\
{\bf g}^{\alpha\beta}&:=&(-g)^{1/2}g^{\alpha\beta},
\end{eqnarray}
and a comma denotes the ordinary derivative. 
By construction, the following conservation laws are satisfied:
\begin{eqnarray}
\left((-g)\left(T^{\alpha\beta}
+t^{\alpha\beta}_{L-L}\right)\right)_{,\beta}=0.
\end{eqnarray}

Suppose that the internal metric around a Kerr black hole is calculated
for a given trajectory of the particle. In terms of the internal metric
we define 
\begin{eqnarray} 
P^a(T,r) &:=& {1\over 16\pi G}
\int_{|X|=r}d^2 S_j H_{L-L}^{ab0j}{}_{,b}\,, 
\label{eq:momentum} \\ 
J^{ij}(T,r) &:=& {1\over 16\pi G}\int_{|X|=r}d^2 S_k 
\left(X^i H^{ja0k}_{L-L}{}_{,a}-X^j H_{L-L}^{ia0k}{}_{,a}\right.
\nonumber\\
&&\hspace{4cm}\left.+H_{L-L}^{ik0j}-H_{L-L}^{jk0i}\right),
\label{eq:spin} 
\end{eqnarray} 
where $d^2 S_j$ is the surface element of a two-sphere at $|X|=r$.
Then by using the Einstein equations (\ref{eq:minoEieq}),
we have the following integral identities:
\begin{eqnarray} 
{d\over dT}P^a(T,r) &=& \int_{|X|=r} d^2 S_j (-g)t_{L-L}^{aj}(X),
\label{eq:force} \\ 
{d\over dT}J^{ij}(T,r) &=& \int_{|X|=r} d^2 S_k 
\left(X^i (-g)t_{L-L}^{jk}(X) -X^j (-g)t_{L-L}^{ik}(X) \right).
\label{eq:torque}
\end{eqnarray}
These are called the laws of motion and precession.
By explicitly evaluating the right hand sides of
Eqs.~(\ref{eq:momentum}), (\ref{eq:spin}), (\ref{eq:force}) and
(\ref{eq:torque}), and eliminating $P^a(T,r)$ and $J^{ij}(T,r)$ from the 
resulting equations, one obtains the equations of motion.


\subsection{Use of the matched asymptotic expansion}

In the present method, we construct the external metric and use the
matching conditions to obtain the necessary components of the internal
metric. The $({}^0_n)$-components of the internal metric are assumed to
be given by the metric of a Kerr black hole.
Since we do not construct the internal metric independently,
there exists no a priori requirement for thus obtained internal
metric to satisfy some specific gauge condition. Hence the
transformation from the external coordinates to the internal ones can be
rather arbitrarily chosen.
Here, we make use of the knowledge we have obtained in section
\ref{sec:deri1} and we choose the coordinate conditions as follows.

We assume that the external metric is generated by the point-like
source, Eq.~(\ref{eq:point}), and calculate the external metric in
the matching region as in the previous section. In order to do so,
the hypersurfaces of $T={\rm constant}$ and $T_x={\rm constant}$ should
be identical to each other as given by Eq.~(\ref{eq:hypcond}).
To satisfy this requirement, we adopt the coordinate transformation
from $x$ to $X$ in the form,
\begin{equation}
\sigma_{;\alpha}(x,z(T)) + f_{\alpha i}(T) X^i = O((Gm)^2/L).
\label{eq:minosurface}
\end{equation}
This is satisfied by setting
\begin{equation}
Lf^\alpha{}_{ij} = L^2f^\alpha{}_{ijk} = O(Gm/L)\,,
\label{eq:fijfijk0}
\end{equation}
in Eq.~(\ref{eq:calFexpand}). Note that, in the case of a monopole
particle discussed in section \ref{sec:deri1}, the conditions that
are required to guarantee Eq.~(\ref{eq:hypcond}) are obtained from the
$({}^{n}_0)$-matchings ($n=0,1,2$). On the contrary, here we impose the
conditions (\ref{eq:fijfijk0}) by hand to guarantee
Eq.~(\ref{eq:hypcond}). 

Furthermore, to determine the internal metric from the matching
conditions, we set the $({}^1_n)$-components of the internal metric to
zero:
\begin{equation}
  \label{eq:H1ncond}
  {}^{(1)}_{(n)}H_{ab}=0\quad (n=0,1,2,\cdots).
\end{equation}
In the case of a monopole particle, we have found we can impose these
conditions. However, in the present case, since we have imposed the
coordinate condition (\ref{eq:minosurface}) by hand, it is not clear if
a similar argument can be made to justify these conditions. Nevertheless,
at least for $n=0$, $1$, we should be able to require the conditions
(\ref{eq:H1ncond}). This is because the spin of the black hole appears
at $O\bigl((Gm)^2\bigr)$ or higher in the internal metric, hence the
discussion we gave in the case of a monopole particle should be equally
applicable to the $({}^1_0)$ and $({}^1_1)$-components of the metric.
In fact, we see below that the conditions (\ref{eq:H1ncond}) for $n=0$,
$1$ consistently determine the internal metric in the local rest frame
by matching. 

First consider the background metric in the internal scheme. 
For convenience we define the trace-reversed $({}^m_n)$-components of
the metric with respect to the flat Minkowski space:
\begin{eqnarray}
{}^{(m)}_{(n)}\bar H_{ab}={}^{(m)}_{(n)}H_{ab}
-{1\over 2}\eta_{ab}\eta^{cd}{}^{(m)}_{(n)}H_{cd}
\end{eqnarray}
Expanding the Kerr metric with respect to $Gm$, the
$({}^0_n)$-components of the metric in the harmonic coordinates are
found as
\begin{eqnarray}
{}^{(0)}_{(0)}H_{ab}&=&\eta_{ab}\,,
\label{eq:H00ab}\\
Gm\,{}^{(0)}_{(1)}\bar H_{TT}&=& {4Gm \over |X|} \,,
\label{eq:H01TT}\\
Gm\,{}^{(0)}_{(1)}\bar H_{Ti}&=& 0 \,,
\label{eq:H01Ti}\\
Gm\,{}^{(0)}_{(1)}\bar H_{ij}&=& 0 \,,
\label{eq:H01ij}\\
(Gm)^2{}^{(0)}_{(2)}\bar H_{TT} &=& {(Gm)^2 \over|X|^2} \,,
\label{eq:H02TT}\\
(Gm)^2{}^{(0)}_{(2)}\bar H_{Ti} &=&{2 G m \over |X|^3}S_{ij}X^j\,, 
\label{eq:H02Ti}\\
(Gm)^2{}^{(0)}_{(2)}\bar H_{ij} &=& {(Gm)^2 \over |X|^2} 
\left(-2 \delta_{ij} +{X^i X^j\over |X|^2}\right) \,,
\label{eq:H02ij}
\end{eqnarray}
where $S_{ij}$ is the specific spin tensor in the local rest frame of
the black hole. Then calculating the $({}^0_0)$ and
$({}^0_1)$-components of the external metric in the matching region, we
find they are consistent with  Eqs.~(\ref{eq:H00ab}) $\sim$
(\ref{eq:H01ij}) provided that $\dot z(T)$ and $f^\alpha_i(T)$ satisfy
the lowest order orthonormal conditions, Eqs.~(\ref{eq:m0tt}),
(\ref{eq:m0ti}) and (\ref{eq:m0ij}). Further, the spin contribution to
the metric, Eq.~(\ref{eq:H02Ti}), can be reproduced from the external
metric with the source (\ref{eq:point}) by the identification,
\begin{equation}
 S_{\alpha\beta}(T)=S_{ij}f_\alpha^i(T) f_\beta^j+O((Gm)^2/L).
\end{equation}
This fact indicates the consistency of using the point-particle energy
momentum tensor (\ref{eq:point}) in the perturbation analysis.

Keeping in mind the imposed conditions (\ref{eq:fijfijk0}) and
(\ref{eq:H1ncond}), the calculation of the $({}^1_0)$-components of the
external metric in the matching region gives
\begin{equation}
  \label{eq:h10ab}
{dz^\alpha\over dT}(T){Df_{\alpha i}\over dT}(T)=O(Gm/L^2),
\quad
f^\alpha_i(T){Df_{\alpha j}\over dT}(T)=O(Gm/L^2).
\end{equation}
As before, these equations imply that $f^\alpha_i(T)$ is
parallel transported along the particle trajectory at the lowest order.
Similarly, the calculation of the $({}^{1}_{1})$-components of the
external metric gives the same conditions as we have found in the
previous section (see Eqs.~(\ref{eq:m1tt}) $\sim$ (\ref{eq:m1ij})):
\begin{eqnarray}
\dot z^2(T) &=& -1 - {Gm \over 2} \bar \Theta_{\alpha\beta}(T) 
\left(g^{\alpha\beta}(z(T)) +2 \dot z^\alpha(T)\dot z^\beta(T)\right) 
\nonumber \\ && \qquad 
+O((Gm)^2/L^2) \,, 
\label{eq:Ltt} \\ 
\dot z^\alpha(T) f_{\alpha i}(T) &=& 
-Gm \bar \Theta_{\alpha\beta}(T) \dot z^\alpha(T) f^\beta{}_i(T) 
\nonumber \\ && \qquad 
+O((Gm)^2/L^2) \,, 
\label{eq:Lti} \\ 
f^\alpha{}_i(T)f_{\alpha j}(T) &=& \delta_{ij} 
-{Gm \over 2} \bar \Theta_{\alpha\beta}(T)\left(-\delta_{ij} 
g^{\alpha\beta}(z(T))
+2f^\alpha{}_i(T)f^\beta{}_j(T)\right)
\nonumber \\ && \qquad 
+O((Gm)^2/L^2) \,, 
\label{eq:Lij} 
\end{eqnarray}
where we have introduced
\begin{eqnarray}
\bar \Theta_{\alpha\beta}=\Theta_{\alpha\beta}
-1/2 g_{\alpha\beta} \Theta^\delta{}_{\delta}
={1\over Gm}\bar\psi_{(tail)\alpha\beta} \,.
\end{eqnarray}
These equations may be viewed as a coordinate condition on the internal
time $T$. Clearly there is no inconsistency in them.

Computation of the rest of components of the internal metric which are
needed to derive the equations of motion is straightforward. The results
are 
\begin{eqnarray}
{1\over L^2}\,{}^{(2)}_{(0)}\bar H_{TT}
&=& - {2\over 3}R_{\alpha\beta\gamma\delta}(z(T)) 
\dot z^\alpha(T) X^\beta(T) \dot z^\gamma(T) X^\delta(T) 
\nonumber \\ && 
+O(Gm|X|^2/L^3) \,,
\\
{1\over L^2}\,{}^{(2)}_{(0)}\bar H_{Ti}
&=& -{2\over 3}R_{\alpha\beta\gamma\delta}(z(T))
f^\alpha{}_i(T) X^\beta(T) \dot z^\gamma(T) X^\delta(T) 
\nonumber \\ && 
+O(Gm|X|^2/L^3) \,,
\\
{1\over L^2}\,{}^{(2)}_{(0)}\bar H_{ij}
&=& -{1\over 3}R_{\alpha\beta\gamma\delta}(z(T))
X^\beta(T) X^\delta(T) \left(f^\alpha{}_i(T) f^\gamma{}_j(T) 
+\delta_{ij}\dot z^\alpha(T) \dot z^\gamma(T)\right) 
\nonumber \\ && 
+O(Gm|X|^2/L^3) \,,
\\
{Gm\over L^2}\,{}^{(2)}_{(1)}\bar H_{TT}&=& 
\dot z^\alpha(T) {D\over dT}f_{\alpha i}(T) X^i(T)
\nonumber \\ && 
-{10 G m \over 3 |X|}R_{\alpha\beta\gamma\delta}(z(T)) 
\dot z^\alpha(T) X^\beta(T) \dot z^\gamma(T) X^\delta(T) 
\nonumber \\ && 
+ Gm \bar \Theta_{\alpha\beta\gamma}(T) \dot z^\alpha(T) \dot z^\beta(T) 
X^\gamma(T) 
+O((Gm)^2 |X|/L^3) \,,
\\
{Gm\over L^2}\,{}^{(2)}_{(1)}\bar H_{Ti}&=& 
f^\alpha{}_i(T) {D\over dT}f_{\alpha j}(T) X^j
\nonumber \\ && 
+ G m R_{\alpha\beta\gamma\delta}(z(T)) X^\beta(T) \dot z^\gamma(T) 
\left(-\dot z^\alpha(T) f^\delta{}_i(T) 
+{2 \over 3 |X|}f^\alpha{}_i(T) X^\delta(T)\right)
\nonumber \\ && 
+ Gm \bar \Theta_{\alpha\beta\gamma}(T)\dot z^\alpha(T) 
f^\beta{}_i(T) X^\gamma(T) 
+O((Gm)^2 |X|/L^3) \,,
\\
{Gm\over L^2}\,{}^{(2)}_{(1)}\bar H_{ij}&=& 
\delta_{ij}\dot z^\alpha(T) {D\over dT}f_{\alpha k}(T) X^k
\nonumber \\ && 
-2 G m R_{\alpha\beta\gamma\delta}(z(T)) 
\Bigl({1\over 3|X|}X^\beta(T) X^\delta(T) 
\left(f^\alpha{}_i(T) f^\gamma{}_j(T) 
-\delta_{ij}\dot z^\alpha(T) \dot z^\gamma(T) \right) 
\nonumber \\ && \qquad \qquad \qquad \qquad \qquad 
-2 |X| \dot z^\alpha(T) f^\beta{}_i(T) \dot z^\gamma(T) f^\delta{}_j(T) 
\Bigr)
\nonumber \\ && 
-2 Gm \Theta_{\alpha\beta\gamma}(T) 
f^\alpha{}_i(T) f^\beta{}_j(T) X^\gamma(T)
+O((Gm)^2 |X|/L^3) \,,
\end{eqnarray}
where $X^\alpha(T)=f^\alpha_i(T)X^i$ and we have defined 
\begin{eqnarray}
\bar \Theta_{\alpha\beta\gamma}:=\Theta_{\alpha\beta\gamma}
-1/2 g_{\alpha\beta} \Theta^\delta{}_{\delta\gamma} \,. 
\end{eqnarray}


\subsection{Equations of motion}

Before evaluating Eqs.~(\ref{eq:momentum}) and (\ref{eq:force}), 
let us first consider the equations for the spin, Eqs.~(\ref{eq:spin})
and (\ref{eq:torque}). Equation (\ref{eq:spin}) has a dimension of
$(mass)\times(length)$ and we extract out the terms of $O(Gm^2)$.
Power counting of $X$ shows that there will be contributions 
linear in the $({}^0_2)$-components of the metric and those from
bilinear combinations of the $({}^0_1)$- and $({}^0_1)$-components
of the metric. Then we obtain 
\begin{eqnarray}
J^{ij}(T,r) = m S_{\alpha\beta}(T)f^{\alpha i}f^{\beta i} +O(G^2m^3/L)
+(r\mbox{-dependent terms}). 
\label{eq:spin1}
\end{eqnarray}
Equation (\ref{eq:torque}) has a dimension of $(mass)^1$ and 
we extract out the terms of $O(Gm^2/L)$ in the same way. 
Power counting of $X$ shows that 
there will be contributions from bilinear combinations of
the $({}^0_1)$- and $({}^0_1)$-components of the metric, 
and we find that the right hand side of Eq.~(\ref{eq:torque}) vanishes:
\begin{eqnarray}
{d \over dT}J^{ij}(T,r) = O(G^2m^3/L^2) 
+(r\mbox{-dependent terms}). 
\label{eq:torque1}
\end{eqnarray}
Since the spatial triad are geodetic parallel transported 
in the background geometry to the leading order, 
Eqs.~(\ref{eq:spin1}) and (\ref{eq:torque1}) result in 
\begin{eqnarray}
{D\over dT}S^{\alpha\beta}(T) =O\left({(Gm)^2\over L^2}\right).
 \label{eq:parallel1}
\end{eqnarray}
Thus in the test particle limit $m\to0$ the spin tensor is parallel
transported along the particle trajectory in the background geometry.

We next consider Eqs.~(\ref{eq:momentum}) and (\ref{eq:force}). 
Equation (\ref{eq:momentum}) has a dimension of $(mass)^1$ and 
we extract out the terms of $O(m)$ and $O(Gm^2/L)$. 
We find that there will be linear contributions 
from $({}^0_1)$-, $({}^0_2)$-components of the metric, 
and bilinear contributions 
from pairs of $({}^0_1)-$ and $({}^0_1)$-components of the metric. 
We obtain 
\begin{eqnarray}
P^0(T,r) &=& m + O(G^2 m^3/L^2) +(r\mbox{-dependent terms}), \\ 
P^i(T,r) &=& O(G^2 m^3/L^2) +(r\mbox{-dependent terms}). 
\end{eqnarray}
Eq.~(\ref{eq:force}) has a dimension of $(mass)/(length)$ and 
we consider the terms of $O(m/L)$ and $O(Gm^2/L^2)$. 
There will be bilinear contributions from 
pairs of the $({}^0_2)-$ and $({}^2_0)$-components 
and pairs of the $({}^0_1)-$ and $({}^2_1)$-components of the metric. 
We find the former pairs give the spin-induced force 
and the latter pairs give the radiation reaction force. 
A straightforward computation results in 
\begin{eqnarray}
{d\over dT}P^0(T,r) &=& O(G^2 m^3/L^3) +(r\mbox{-dependent terms}), 
\\ 
{d\over dT}P^i(T,r) &=& 
-{m\over 2}R_{\alpha\beta\gamma\delta}(z(T))
f^\alpha{}_i(T)\dot z^\beta(T)S^{\gamma\delta}(T)
\nonumber \\ && 
+{Gm^2 \over 4} \bar \Theta_{\alpha\beta\gamma}(T)f^\gamma{}_i(T)
\left(2\dot z^\alpha(T)\dot z^\beta(T)+g^{\alpha\beta}(z(T))\right)
\nonumber \\ && 
+m\dot z^\alpha(T) {D\over dT}f_{\alpha i}(T) 
\nonumber\\&&
+O(G^2 m^3/L^3) +(r\mbox{-dependent terms}). 
\end{eqnarray}
Taking the $T$-derivative of Eqs.~(\ref{eq:Ltt}) and (\ref{eq:Lti}), 
we obtain the equations of motion, 
\begin{eqnarray}
{D \over dT}\dot z^\alpha(T) 
&=& -{Gm \over 2} \Theta_{\beta\gamma\delta}(T) 
\left(2\dot z^\beta(T)g^{\alpha\gamma}(z(T))\dot z^\delta(T) 
-\dot z^\beta(T)\dot z^\gamma(T)g^{\alpha\delta}(z(T))\right) 
\nonumber \\ && 
-{1\over 2}R^\alpha{}_{\beta\gamma\delta}(z(T))
\dot z^\beta(T)S^{\gamma\delta}(T)
+O(G^2 m^2/L^3) \,. \label{eq:result0} 
\end{eqnarray}
Introducing the proper time $\tau$ of the orbit, 
\begin{eqnarray}
{d\tau \over dT}&=& 
1+  {Gm \over 2}\bar \Theta_{\alpha\beta}(T)\dot z^\alpha(T)\dot z^\beta(T), 
\end{eqnarray}
we finally arrive at
\begin{eqnarray}
{D \over d\tau}{d z^\alpha \over d\tau}(\tau) &=& 
-{Gm \over 2}\Theta_{\beta\gamma\delta}(\tau) 
\left({d z^\alpha \over d\tau}{d z^\beta \over d\tau}
{d z^\gamma \over d\tau}{d z^\delta \over d\tau}
+2{d z^\beta\over d\tau}g^{\alpha\gamma}(\tau){d z^\delta\over d\tau} 
-{d z^\beta\over d\tau}{d z^\gamma\over d\tau}g^{\alpha\delta}(\tau)
\right) 
\nonumber \\ && 
-{1\over 2}R^\alpha{}_{\beta\gamma\delta}(\tau)
{d z^\beta\over d\tau}S^{\gamma\delta}(\tau)
+O(G^2 m^2/L^3) \,,
\label{eq:result}
\end{eqnarray}
where $Q(\tau)=Q(z(\tau))$. One finds that the result is exactly equal to
Eq.~(\ref{eq:damp1r}) except for the spin-curvature coupling term.

In the case of a monopole particle discussed in the previous
section, the $({}^2_1)$ matching gave two conditions 
(\ref{eq:m3/2ttg}) and (\ref{eq:m3/2tig}). 
The latter condition was crucial to obtain the $O(Gm/L^2)$ correction
terms in the evolution equations of $f^{\alpha}_i$.
In the present analysis, we do not have the counterpart of this
condition. This indicates that the gauge condition relating with the
notational mode must be specified to determine $Df_{\alpha i}/dT$.


\section{Discussion}

Let us first discuss the physical meaning of the equations of motion
obtained in the preceding two sections.  
For simplicity, we consider the case of a monopole particle.
We divide the perturbed metric in the external scheme  
into the two:
\begin{equation}
h_{\mu\nu}(x)=h_{(mono)\mu\nu}(x)+h_{(tail)\mu\nu}(x), 
\end{equation}
where $h_{(tail)\mu\nu}(x)$ is the part due to the
$v^{\mu\nu\alpha\beta}$ in the Green function while 
$h_{(mono)\mu\nu}$ is due to the $u_{\mu\nu\alpha\beta}$ term
(see Eq.~(\ref{eq:metper})). 
The singular behavior of the perturbed metric 
in the coincidence limit $x\to z$ is totally due to
$h_{(mono)\mu\nu}(x)$. Thus, we introduce the regularized 
perturbed metric as
\begin{eqnarray}
\tilde g_{(reg)\mu\nu}(x):=g_{\mu\nu}(x)+h_{(tail)\mu\nu}(x),
\label{eq:regmet}
\end{eqnarray}
which has no singular behavior any more. 
Then we find the equations of motion (\ref{eq:damp1}) and the 
evolution equations of the triad basis (\ref{eq:damp2}) 
coincide with the geodesic equation and the geodetic parallel transport
equation, respectively, on the regularized spacetime with
the metric $\tilde g_{(reg)\mu\nu}$. 
To see this let us consider the parallel transport of a vector 
$A^{\alpha}$ along a geodesic $x^\alpha=z^\alpha(\tilde\tau)$
in this spacetime. It is given by
\begin{equation}
 {\tilde D\over d\tilde\tau}A^{\alpha}:=
  {D\over d\tilde\tau}A^{\alpha}+\delta 
  \Gamma_{(reg)}{}^{\alpha}{}_{\beta\gamma} A^{\beta}
  {dz^{\gamma}\over d\tilde\tau}=0, 
\label{eq:saigo}
\end{equation}
to the linear order in $h_{(tail)\mu\nu}$ where 
\begin{equation}
  \delta \Gamma_{(reg)}{}^{\alpha}{}_{\beta\gamma}:=
   {1\over 2}\left(h_{(tail)}{}^{\alpha}{}_{\beta;\gamma}+
   h_{(tail)}{}^{\alpha}{}_{\gamma;\beta}-h_{(tail)\beta\gamma}{}^{;\alpha}
   \right) \,. 
\end{equation}
Then one recovers Eqs.~(\ref{eq:damp1}) and (\ref{eq:damp2}) by 
identifying $\tilde\tau$ with $T$ and 
replacing $A^\alpha$ with $dz^\alpha/d T$ or $f^{\alpha}{}_{i}$.
In the case of a spinning particle, 
there exists an additional force in the equations of motion 
(\ref{eq:result0}) due to the coupling of 
the spin and the background curvature. 

The result for the monopole particle seems analogous to that in 
the electromagnetic case\cite{DeWitt}, 
except that the instantaneous reaction force which is proportional to
higher derivatives of the particle velocity is absent in the present
case. This is because the particle traces a geodesic in the lowest order
approximation.  If an external force field exists, the assumption of the 
geodetic motion in the lowest order breaks down and furthermore 
the contribution of the external force field 
to the energy momentum tensor must be taken into account.
Since this fact makes the problem too complicated,
it is beyond the scope of the present discussion. 

Now let us consider how to construct $\tilde g_{(reg)\mu\nu}$, 
in the case of a monopole particle. 
Unfortunately, we do not have any satisfactory formalism that can be
applied to such a calculation, even for a specific background spacetime
such as the Kerr geometry, mainly due to the difficulty in evaluating
the bi-tensor $v_{\mu\nu\alpha\beta}(x,z)$.
Here we just give a few primitive discussions on this matter. 

Basically, there seems to be two approaches
for calculating $\tilde g_{(reg)\mu\nu}$ (or equivalently 
$h_{(tail)\mu\nu}$). 
The first one is to calculate $h_{(tail)\mu\nu}$ directly. 
The second one is to calculate 
$h_{\mu\nu}=h_{(mono)\mu\nu}+h_{(tail)\mu\nu}$ and subtract 
$h_{(mono)\mu\nu}$ from it. 
In the following, we discuss only the first approach. 
As for the second approach, we have nothing to mention here, but
this direction of research may be fruitful\cite{Minothesis}. 

By definition, $h_{(mono)\mu\nu}$ evaluated on the particle trajectory 
is independent of the past history of the particle.\footnote{
There is a possibility that 
the future light cone emanating from $z$ crosses 
the particle trajectory again. Since inclusion of this possibility
 makes the problem too complicated, we do not consider it here.}
Therefore if we consider the metric defined by 
\begin{equation}
 h^{(\Delta\tau)}_{\mu\nu}(x)=
  G m \left(\delta_{\mu}{}^{\rho}\delta_{\nu}{}^{\sigma}
    -{1\over 2}g_{\mu\nu}(x) g^{\rho\sigma}(x)\right)
   \int_{-\infty}^{\tau_x-\Delta\tau}~
  d\tau' G^{Ret}_{\rho\sigma\alpha\beta}(x,z(\tau'))\dot z^{\alpha}
   (\tau')\dot z^{\beta}(\tau'), 
\end{equation}
for any finite $\Delta\tau$\,$(>0)$, it will not contain 
$h_{(mono)\mu\nu}$ when it is evaluated on the particle trajectory. 
The difference between $h^{(\Delta\tau)}_{\mu\nu}$ and 
$h_{(tail)\mu\nu}$ comes from the integral 
over a small interval, 
\begin{equation}
   \sim Gm \int_{\tau_x-\Delta\tau}^{\tau_x}~
  d\tau' v_{\rho\sigma\alpha\beta}(x,z(\tau'))\dot z^{\alpha}
   (\tau')\dot z^{\beta}(\tau').
\end{equation}
Since $v_{\rho\sigma\alpha\beta}(x,z)$ is regular in the coincidence
limit $x\to z$, this integral will be negligible for a sufficiently
small $\Delta\tau$. Thus
$\lim_{\Delta\tau\to0}h^{(\Delta\tau)}_{\mu\nu}$ will give 
$h_{(tail)\mu\nu}$.

In the case of the electromagnetic (vector) Green function,
a calculation along the above strategy was performed by
DeWitt and DeWitt\cite{DeW2} by assuming
the background gravitational field is weak so that
its metric is given by the small perturbation 
on the Minkowski metric,
\begin{equation}
 g_{\mu\nu}=\eta_{\mu\nu}+h^{(b)}_{\mu\nu}\,.
\end{equation}
DeWitt and DeWitt calculated the relevant part of the
Green function perturbatively to the first order in $h^{(b)}_{\mu\nu}$
by using the Minkowski Green function.

Here we should mention one important fact. We have obtained the
equations of motion with the correction term of $O(Gm/L^2)$.
Although we use the terminology `radiation reaction' to describe it, it
is not appropriate in a narrow sense because the correction term may
well contain something more than just the usual effect of radiation
reaction. In fact, in the electromagnetic case, the existence of 
the effect which is termed as `the induced polarization force on the
background spacetime' is reported by several authors\cite{inpol}. 
Furthermore, a calculation analogous to that done by DeWitt and
DeWitt\cite{DeW2} for the electromagnetic case was done by 
Carmeli\cite{Carm} for the gravitational case and it was shown that the
tail part correctly reproduces the lowest order post-Newtonian
corrections to the equations of motion. However, no such calculation has
been done for the background with strong gravity, such as 
a black hole spacetime. It seems difficult to develop DeWitt and
DeWitt's approach to higher orders in $h^{(b)}_{\mu\nu}$.
It is a challenging issue to formulate a systematic method to evaluate
the tail part of the metric when the background gravity is strong and
clarify its physical content.

Turning back to the effect of the gravitational radiation reaction,
we should make one additional comment.
There has been some proposals to obtain the radiation reaction force in
a quite different manner. Among others is the use of the radiative Green
function (a half of the retarded minus advanced Green functions) in the
case of a Kerr background proposed by Gal`tsov\cite{Gal}. As easily seen
from the results in section \ref{sec:minoExt}, the use of the
radiative Green function instead of the retarded one results in the
replacement of $\psi_{(tail)\mu\nu}(x)$ by $\psi^{Rad}_{(v)\mu\nu}(x)$,
which is defined by 
\begin{equation}
\psi^{Rad}_{(v)}{}_{\beta\gamma}(x):=-Gm\int^{+\infty}_{-\infty}d\tau' 
v_{\beta\gamma\alpha'\beta'}(x,z(\tau'))
\dot z^{\alpha'}(\tau')\dot z^{\beta'}(\tau'). 
\end{equation}
Gal'tsov proved that the back reaction force computed using the
radiative Green function gives the loss rates of the energy and the
$z$-component of the angular momentum of the particle in quasi-periodic
orbits which correctly balance with the emission rates of the
corresponding quantities by gravitational radiation.
However, we do not think that this fact indicates 
the correctness of the prescription, even if we restrict it to the case
of a Kerr background, because those constants of motion
are special ones which reflect the existence of the corresponding 
Killing vector fields. For such quantities, there may be some symmetry
in the structure of the Green function which makes the use of the
radiative Green function valid. However, it is doubtful that the 
radiative Green function correctly describes the radiation reaction
effect on the Carter constant.

Finally we make a couple of comments on the implications of our results.
It is important to note that the particle does not have to be a black
hole since the detailed internal boundary condition was not used to
determine the metric in the internal scheme. The resulting equations of
motion should be equally applicable to any compact body such as a
neutron star. The essential assumption here is that the only length
scale associated with the particle is $Gm$. 
In this sense, we have shown the strong equivalence principle to
the first order in $Gm$.

We also note that our results strongly support, if not rigorously
justify, the so-called black hole perturbation approach.
In the black hole perturbation approach, one calculates the
gravitational radiation from a particle orbiting a black hole with the
assumption that the particle is a point-like object with the energy
momentum tensor described by the delta function. Although this approach
has been fruitful, there has been always skepticism about the validity
of the delta functional source. What we have shown in this chapter is 
the consistency of using the delta function in the source energy
momentum tensor within the order of matched asymptotic expansion we have 
examined. 


\end{document}